\pdfoutput=1
\documentclass[11pt]{article}
\usepackage[T1]{fontenc}
\usepackage[nott]{newtxtext}

\usepackage[margin=1in]{geometry}
\usepackage{microtype}

\usepackage{mathtools}
\usepackage{amssymb}
\usepackage{amsthm}
\usepackage{newtxmath}
\usepackage{bbm}
\DeclareMathSizes{8}{8}{6}{5}
\DeclareMathSizes{9}{9}{6}{5}
\DeclareMathSizes{10}{10}{7}{5}
\usepackage{thm-restate}
\usepackage[ruled,vlined,linesnumbered,algo2e]{algorithm2e}
\usepackage[inline]{enumitem}

\usepackage[round,authoryear]{natbib}
\usepackage{xcolor}
\usepackage[pdfusetitle]{hyperref}
\usepackage{bookmark}
\definecolor{linkblue}{HTML}{1F5A94}
\definecolor{citegreen}{HTML}{276749}
\hypersetup{
  colorlinks=true,
  linkcolor=linkblue,
  citecolor=citegreen,
  urlcolor=linkblue,
  bookmarksopen=true,
  bookmarksnumbered=true
}
\setlist{nosep}
\setlist[itemize]{leftmargin=1.5em}
\setlist[enumerate]{leftmargin=1.7em}
\allowdisplaybreaks

\theoremstyle{plain}

\newtheorem{propositionn}{Proposition}
\theoremstyle{definition}
\newtheorem{definition}{Definition}
\newenvironment{proofof}[1]{\begin{proof}[Proof of #1]\leavevmode}{\end{proof}}

\newenvironment{protocol}[1][t]{%
  \SetAlgorithmName{Protocol}{protocol}{List of Protocols}%
  \begin{algorithm2e}[#1]%
}{%
  \end{algorithm2e}%
}

\def \A {\mathcal{A}}
\def \B {\mathcal{B}}
\def \E {\mathcal{E}}
\def \F {\mathcal{F}}
\def \G {\mathcal{G}}
\def \L {\mathcal{L}}
\def \O {\mathcal{O}}
\def \T {\mathcal{T}}
\def \Z {\mathcal{Z}}

\def \eps {\epsilon}
\def \del {\delta}
\def \lam {\lambda}
\def \nat {\mathbb{N}}
\def \rr {\mathbb{R}}
\def \zz {\mathbb{Z}}
\def \poly {\textup{poly}}
\DeclareMathOperator*{\expec}{\mathbb{E}}
\DeclareMathOperator*{\argmin}{arg\,min}
\newcommand{\innp}[1]{\left\langle #1 \right\rangle}
\newcommand{\cbra}[1]{\left\{ #1 \right\}}
\newcommand{\sbra}[1]{\left[ #1 \right]}
\newcommand{\ceil}[1]{\left\lceil #1 \right\rceil}
\newcommand{\floor}[1]{\left\lfloor #1 \right\rfloor}
\newcommand{\ind}{\mathbbm{1}}
\newcommand{\bunderbrace}[2]{%
  \begin{array}[t]{@{}c@{}}
  \underbrace{#1}\\
  #2
  \end{array}
}
\newcommand\numberthis{\addtocounter{equation}{1}\tag{\theequation}}

\def \ber {\mathsf{Ber}}
\def \reg {\mathsf{Reg}}
\def \sreg {\mathsf{SReg}}
\def \sm {\mathsf{SUM}}
\def \opt {\mathsf{Opt}}
\def \rev {\mathsf{Rev}}
\def \irev {\mathsf{rev}}
\def \nash {\mathsf{NE}}
\def \omr {\mathsf{OMR}}
\def \omrgame {\omr(\A,\B,\E,\T)}
\def \omrgameTruthful {\omr(\A^\L_\omr,\Btru,\E,\T)}
\def \omrgameSR {\omr(\A^\L_\sm,\B,\E,\T)}
\def \Btru {\B_{\textup{tru}}}
\def \bargam {\overline{\gamma}}
\def \bb {\mathbb{B}}
\def \sign {\mathsf{sign}}

\title{Strategy-robust Online Learning in Contextual Pricing}
\author{
  Joon Suk Huh \qquad Kirthevasan Kandasamy\\[0.4em]
  \small University of Wisconsin--Madison\\
  \small\texttt{joon@cs.wisc.edu}\qquad
  \small\texttt{kandasamy@cs.wisc.edu}
}
\date{}
\hypersetup{pdfauthor={Joon Suk Huh and Kirthevasan Kandasamy}}

\begin{document}
\maketitle

\begin{abstract}
Learning effective pricing strategies is crucial in digital marketplaces, especially when buyers’ valuations are unknown and must be inferred through interaction.
We study online contextual pricing with sealed bids from multiple recurring buyers. The seller selects an affine pricing policy before observing the current context and competes with the best fixed policy in the same class. Buyers may misreport their values to influence future prices, a challenge known as strategic overfitting~\citep{amin2013learning}.

We introduce a strategy-robust notion of regret for multi-buyer online environments, capturing worst-case strategic behavior in the spirit of the Price of Anarchy.
Our first contribution is a polynomial-time approximation scheme (PTAS) for learning affine pricing policies against arbitrary context--value sequences: for every fixed $\eps>0$, it runs in polynomial time and its regret is $\O(\eps T)$ plus a term sublinear in $T$. The scheme is based on an online sketch constructed from past contexts. We then propose the Sparse Update Mechanism (SUM), which learns only from a privately selected subset of rounds and occasionally posts a random price drawn from a finite grid. SUM is a PTAS for strategic regret, and its guarantee holds uniformly over all Nash equilibria. Both schemes use an online expert algorithm as a black box and run in $(T/\eps)^{\O(\eps^{-2})}\poly(d,T,1/\eps)$ time. Furthermore, we show that, even with truthful bids, no seller algorithm can achieve regret $\poly(d,1/\eps)\,T^z+\eps T$ with $z<1$ in time $\poly(d,T,1/\eps)$ unless $\mathrm{NP}\subseteq\mathrm{BPP}$.
\end{abstract}

\noindent\textbf{Keywords:}
Online Learning; Dynamic Pricing; Mechanism Design; Algorithmic Game Theory

\section{Introduction}\label{sec:intro}
The ability to learn how to price effectively is a key component of modern digital markets, enabling the efficient exchange of goods and services. This problem has been studied extensively~\citep{gallego1994optimal,kleinberg2003value,besbes2009dynamic,dasu2010dynamic,adamczyk2015sequential,besbes2015surprising,den2015dynamic,vanunts2019optimal,leme2021learning,romano2021online,jazi2025posted} since the rise of e-commerce. When the seller knows the distribution of buyers' valuations for a good, the revenue-maximizing price can be computed directly from that distribution. In the learning-to-price framework, however, the seller does not know the value distribution and must instead learn it from data on buyers’ valuations.

In many practical settings, pricing decisions benefit from leveraging not only valuation data but also contextual features that capture relevant information about the goods, buyers, or the market. This gives rise to the framework of \textit{contextual pricing}, where machine learning models are used to map contexts to prices. In the \textit{online contextual pricing} setting, the seller observes a stream of context–valuation pairs and learns a pricing policy sequentially. In this work, the price is the positive part of an affine function of the context—a simple yet expressive model that underlies more complex approaches such as kernel methods and neural networks. The seller privately selects its policy before observing the current context and competes with the best fixed policy of this form. The corresponding batch setting was formalized as Myersonian regression~\citep{mohri2014learning, munoz2017revenue,shen2019learning,liu2020myersonian}, and our work extends its ideas to the online setting, accounting for recurring strategic bidders with discounted utilities.

Most prior work assumes access to a reliable data source, either in batch~\citep{mohri2014learning, munoz2017revenue,shen2019learning,liu2020myersonian} or online form~\citep{javanmard2019dynamic,tang2020differentially,ban2021personalized,den2022dynamic,tullii2024improved}, that accurately reflects buyers’ true valuations. This assumption is reasonable in the batch setting, where data can be collected through market research or other trusted channels. In contrast, the assumption is much more fragile in the online setting, where data is generated through sequential interactions with buyers. Here, the seller typically elicits a buyer’s valuation through a bidding mechanism, effectively turning the process into a form of \emph{repeated auction}\footnote{Often called the online posted-price auction~\citep{kleinberg2003value}.}. This opens the door to strategic behavior. There is no \textit{a priori} reason to believe that buyers will report their valuations truthfully or refrain from manipulating the pricing algorithm to their own benefit. The problem becomes especially pronounced when buyers interact with the seller multiple times, rather than appearing only once.

As a concrete example, consider a buyer, such as a bakery, that purchases milk daily from a producer. The buyer could strategically underbid—offering less than their true valuation—for several days, resulting in no transactions. This behavior may send misleading signals to the producer, potentially causing them to lower the price in subsequent days. As a result, the buyer may make future purchases at a significantly reduced cost, ultimately increasing the buyer's utility at the expense of the producer’s revenue, despite earlier utility losses due to declined transactions.

\paragraph{Strategy-robustness.}
A line of work~\citep{amin2013learning,amin2014repeated,mohri2015revenue,deng2019robust, drutsa2020optimal, zhiyanov2020bisection, liu2024contextual, golrezaei2023incentive,liu2018learning}, beginning with~\citet{amin2013learning}, has studied the challenge of learning under strategic behavior, with a focus on preventing the phenomenon of \textit{strategic overfitting}, as illustrated in the example above. These works establish regret bounds under the assumption that buyers act to maximize their own surplus. A natural question in this context is whether a regret guarantee applies to a particular, potentially best-case, utility-maximizing behavior, or whether the guarantee is robust to \emph{all} possible utility-maximizing behaviors. In game-theoretic terms, this distinction corresponds to two different notions: the \textit{Price-of-Stability} (PoS)~\citep{schulz2003performance}, which measures efficiency under best-case strategic behavior, and the \textit{Price-of-Anarchy} (PoA)~\citep{koutsoupias1999worst,chung2008price,dubey1986inefficiency}, which captures efficiency under \emph{worst-case strategic behavior}. Hence, we refer to PoA-type guarantees as \emph{strategy-robust}.

In this work, we introduce a natural multi-buyer generalization of regret that explicitly captures the notion of Price-of-Anarchy, which we refer to as \emph{strategy-robust regret}. That is, our regret guarantee holds even under worst-case utility-maximizing strategic behavior across all buyers. Importantly, because our setting is inherently multi-buyer, the notion of a ``utility-maximizing buyer'' is replaced by that of a Nash equilibrium, making our formulation more directly aligned with the concept of Price-of-Anarchy. To the best of our knowledge, this is the first result to establish a PoA-style regret bound in the multi-agent contextual pricing setting.
Moreover, we achieve this form of strategy robustness in a black-box manner, by treating any no-regret online expert algorithm as an oracle, without relying on sophisticated primitives such as differential privacy~\citep{dwork2014algorithmic,liu2018learning,abernethy2019learning,huh2024nash,zhang2021more}.

\paragraph{Contributions and Organization.}
In \S\ref{sec:prelim}, we define the multi-buyer problem, \emph{Online Myersonian Regression} (OMR), and its strategy-robust regret~\eqref{eq:Sreg}. The benchmark is the best fixed affine pricing policy evaluated on the true values, and the regret is measured at the worst Nash equilibrium of the buyers. The true context--value sequence may be arbitrary and fixed in advance, while buyers' bids remain adaptive. We also allow environments that choose the next context and value from past public observations, but not from private bids or the seller's current policy.

We complement this problem setting with a complexity-theoretic barrier: we show that no-regret learning in OMR is computationally hard even when all buyers are truthful. Specifically, unless $\mathrm{NP}\subseteq\mathrm{BPP}$, no polynomial-time seller restricted to affine pricing policies guarantees expected standard regret $C(d,\ell)T^z$ for a fixed $z<1$ and a polynomial $C$, even on IID instances. Here, $\ell$ bounds the bit length of each observation. The proof builds on the hardness of proper halfspace learning~\citep{guruswami2009hardness}, in which the learner must output a halfspace, and reduces this problem to affine pricing.

In \S\ref{sec:truthful}, we address the first challenge: learning from adaptive inputs, assuming truthful bids. In batch Myersonian regression, a fixed Johnson--Lindenstrauss projection reduces the search space~\citep{johnson1984extensions,liu2020myersonian}. Such nonadaptive sketches are not guaranteed to work on adaptive inputs~\citep{hardt2013how}. Our inputs are adaptive: a permitted environment may react to past prices, and the strategic bids of \S\ref{sec:multibuyer} depend on past prices even when the true context--value sequence is oblivious. We therefore construct an online sketch (Lemma~\ref{lem:OnlineSketch}): a finite set of experts whose weight vectors are built from past contexts. The construction is related to online covering methods~\citep{zhang2002covering,rakhlin2015online,ben2009agnostic,hardt2010multiplicative}. For any fixed weight vector and context sequence, some expert approximates that vector's inner products with the contexts to within a given accuracy $\eps>0$ on all but $\O(\eps^{-2})$ rounds. We account for the revenue lost on these exceptional rounds separately.

Combining this sketch with a grid of bias values gives a polynomial-time approximation scheme (PTAS) for affine pricing that uses an online expert algorithm as a black box. With Hedge, Alg.~\ref{alg:omr} has regret
$\O(\eps^{-1}\sqrt{T\log(T/\eps)}+\eps T)$
and running time $(T/\eps)^{\O(\eps^{-2})}\poly(d,T,1/\eps)$
(Theorem~\ref{thm:RegretOMR}). This running time is polynomial for each fixed $\eps$ but not in $1/\eps$. The barrier above explains why: no seller algorithm can achieve regret $\poly(d,1/\eps)\,T^z+\eps T$ with $z<1$ in time $\poly(d,T,1/\eps)$ unless $\mathrm{NP}\subseteq\mathrm{BPP}$ (\S\ref{sec:prelim}). On IID data, we return one of the policies used during training, chosen uniformly at random. This policy belongs to the same class, and its expected revenue on a fresh draw falls short of the best policy's by at most the average regret.

In \S\ref{sec:multibuyer}, we turn to the second challenge, strategic buyers, and extend this black-box construction with the \textit{Sparse Update Mechanism} (SUM). The seller privately selects which bids are used for learning, which limits the future influence of any single bid. With probability $\eps$, the seller explores by posting a uniformly random price, which is a constant policy in the same class. Such random prices make bids far from the true value costly in the current round. Together, sparse updates and random prices ensure that bids are approximately truthful in every Nash equilibrium (Lemma~\ref{lem:SparseUpdateNash}). Quantizing the bids makes the mechanism finite, so an equilibrium exists.

The strategic regret with Hedge is
\[
\O\bigg(
 \bigg(\eps^{-1}+\eps^{-5/2}\sqrt{\frac{\bargam}{1-\bargam}}\bigg)
 \sqrt{T\log(T/\eps)}+\eps T
\bigg),
\]
where $\bargam<1$ is an upper bound on the buyers' discount factors (Theorem~\ref{thm:SRegretOMR}). The running time has the same dependence on $d,T,\eps$ as the truthful algorithm.

\paragraph{Related Work.}
We provide here a concise overview of the most relevant prior work. For a more comprehensive discussion, we refer the reader to Appendix~\ref{app:relatedwork}.

A foundational contribution to strategy-robust learning in pricing is due to~\citet{amin2013learning}, who introduced the notion of \emph{strategic regret}, the gap between the seller’s revenue under truthful buyers and that under utility-maximizing, strategic buyers. They proposed an algorithm achieving sublinear strategic regret when buyer valuations are drawn \emph{independently} from an unknown distribution. This work was extended by~\citet{mohri2015revenue}, who provided guarantees against a broader class of strategic buyers that approximately maximize their utility.

The contextual setting, where the seller observes side information (e.g., user features) at the time of sale, was first studied with strategic buyers by~\citet{amin2014repeated}, who introduced strategic regret in this setting. Subsequent work~\citep{deng2019robust, drutsa2020optimal, zhiyanov2020bisection, liu2024contextual, golrezaei2023incentive} built on this framework by refining strategic regret guarantees and extending the model to incorporate additional complexities such as noisy feedback. All of these results operate under a stochastic environment, where buyer values and contexts are drawn from a fixed but unknown distribution.

A separate line of work addresses adversarial environments, where buyer valuations are chosen adaptively. The seminal work of~\citet{liu2018learning} introduced the first algorithm with no strategic regret in this setting, using differential privacy to limit the influence of strategic buyers. In contrast, our approach achieves a similar effect through a simpler mechanism: the seller feeds the reward vector to an external online expert algorithm only on independently sampled rounds. Because our method is agnostic to the internal structure of the expert algorithm, it offers simplicity and broad applicability.

It is worth noting that while strategy-robust learning in pricing frameworks has included multi-buyer settings in seminal works such as~\citet{amin2013learning} and~\citet{liu2018learning}, the regret guarantees in these works are established against individual buyers who act to maximize their own utility. These behaviors may correspond to specific Nash equilibria, but not necessarily to the entire set of equilibria. In particular, it is unclear whether their algorithmic techniques provide guarantees that hold \emph{uniformly} over all possible Nash equilibria.

\section{Preliminaries}\label{sec:prelim}
We begin by formulating the core problem, \emph{Online Myersonian Regression} (OMR), and then define the notion of an online expert algorithm, which will serve as a black-box component in our algorithms.

\paragraph{Notation.}
We use $[n] \coloneq \{1, \dots, n\}$ to denote the set of integers from $1$ to $n$. We set $[0]\coloneq\emptyset$ and take empty sums to be zero. The Euclidean norm is denoted by $\|\cdot\|$, and the inner product by $\innp{v, w}$. The $d$-dimensional Euclidean unit ball is $\bb^d \coloneq \{x \in \rr^d : \|x\| \leq 1\}$. The indicator function is written as $\ind[\cdot]$. For a scalar $q\in\rr$, we write $[q]_+\coloneq\max\{0,q\}$ for its positive part. For a sequence $(x_1, \dots, x_t)$, we write $x_{\leq t} \coloneq (x_1, \dots, x_t)$ and $x_{<t}\coloneq(x_1,\dots,x_{t-1})$. The set of probability distributions over the elements in $S$ is denoted by $\Delta_S$, and $U_S$ denotes the uniform distribution over $S$. The Bernoulli distribution with success probability $p$ is denoted by $\ber(p)$. The notation $\poly(n)$ refers to a polynomial function in $n$.

\subsection{Problem Setting: Online Myersonian Regression}\label{subsec:omr}
The main problem we aim to solve is the online (i.e., sequential) variant of Myersonian regression, introduced by~\citet{liu2020myersonian}, which we refer to as \emph{Online Myersonian Regression} (OMR).
In OMR, a seller $\A$ interacts with a sequence of buyers, selling one good per round. In each round $t \in [T]$, a buyer arrives with a private value $\theta_t \in [0,1]$ for the good. We assume there are $n$ buyers, indexed by $i \in [n]$, who appear over $T$ rounds. Specifically, let $\T_i \subseteq [T]$ denote the set of rounds in which buyer $i$ participates. These sets satisfy $\bigcup_{i \in [n]} \T_i = [T]$ and $\T_i \cap \T_j = \emptyset$ for all $i \ne j$; that is, the collection $\T\coloneq (\T_i)_{i \in [n]}$ forms a partition of $[T]$.

In each round, a normalized context vector $x_t\in\rr^d$ with $\|x_t\|=1$ describes the buyer or the good. Before observing $x_t$, the seller privately selects an \emph{affine pricing policy} from
\begin{align*}
    \F_{\textup{aff}}\coloneq
    \cbra{f_{w,a}:x\mapsto[\innp{w,x}+a]_+:
        w\in\bb^d,\ a\in\rr}.
    \numberthis\label{eq:affineclass}
\end{align*}
The price in round $t$ is $p_t=f_{w_t,a_t}(x_t)$. We do not cap prices at $1$: a price above $1$ results in no sale. The benchmark, defined below, uses the same class.

The key features of OMR are: (1) the context--value pairs are chosen by an \emph{oblivious adversary}, referred to as the environment $\E$, which fixes an arbitrary sequence $(x_t,\theta_t)_{t=1}^T$ before the interaction begins; and (2) buyers' values are elicited through bids, which may depend adaptively on past interactions. Thus, even for an oblivious true context--value sequence, the bids, and hence the rewards that our algorithms learn from, need not be oblivious. Our analysis also allows a deterministic environment that selects the next pair using past public contexts, prices, and any policy coefficients revealed after bidding, but not private bids, allocations, payments, or the update coins of SUM (\S\ref{sec:multibuyer}). The current pair cannot depend on the seller's current private policy. We call these environments and the oblivious ones \emph{permitted}. Protocol~\ref{prot:omr} specifies the interaction.

\paragraph{Buyer Algorithm.}
Each buyer $i\in[n]$ employs a bidding algorithm $\B_i$. In each round $t\in\T_i$, it maps its information history, the current public context, and its current private value to a distribution over bids $\widetilde b_t\in[0,1]$. The mechanism uses the quantized bid
\begin{align*}
    b_t\coloneq Q_h(\widetilde b_t),\qquad
    Q_h(q)\coloneq h\floor{q/h},\qquad
    \G_h\coloneq\{0,h,\dots,1\},\quad h=1/m,
    \numberthis\label{eq:quantizedbid}
\end{align*}
for an integer $m\geq1$, which we call the bid precision. In particular, $Q_h(1)=1$. Allocation and every part of the seller's state that affects future prices depend only on $b_t$. Our algorithms set the bid precision in~\eqref{eq:gridprecision}.

\paragraph{Information Structure.}
The seller is oblivious to $\B$, $\E$, and the partition $\T$. Before bidding, all parties observe $x_t$, and the current buyer observes $\theta_t$. The seller's current policy and price remain private until the bid is submitted. Afterwards, $p_t$ is public; the seller may also reveal $(w_t,a_t)$. Bids, allocations, and payments are private to the seller and the current buyer. The seller does not observe true values, whereas buyers may observe other buyers' values. Each party also observes its own generated variables and remembers its information history.

\begin{protocol}[t]
\caption{$\omrgame$: (Multi-buyer) Online Myersonian Regression}\label{prot:omr}
\DontPrintSemicolon
\textbf{Players:} Seller's algorithm $\A$ and buyers' algorithms $\B\coloneq(\B_i)_{i\in[n]}$.\\
\textbf{Parameters:} Environment $\E$; horizon $T$; partition $\T$; bid precision $m$, with $h=1/m$.\\
\textbf{Constraints:} $T$ and $h$ are public; $\A$ is oblivious to $\B$, $\E$, and $\T$.\\
\vspace{0.25em}
\For{$t\in[T]$}{
    Let $i_t$ be the buyer with $t\in\T_{i_t}$.\\
    $\E$ selects $(x_t,\theta_t)$, with $\|x_t\|=1$ and $\theta_t\in[0,1]$.
    Simultaneously, $\A$ privately selects $(w_t,a_t)\in\bb^d\times\rr$.\\
    Reveal $x_t$ to all parties and $\theta_t$ to buyer $i_t$.\\
    $\B_{i_t}$ submits a sealed bid $\widetilde b_t$; set $b_t\gets Q_h(\widetilde b_t)$.\\
    $\A$ posts $p_t\gets[\innp{w_t,x_t}+a_t]_+$ and may reveal $(w_t,a_t)$.\\
    Sell at $p_t$ if and only if $p_t\leq b_t$.
    Other buyers and $\E$ do not observe the bid, allocation, or payment.
}
\end{protocol}

\paragraph{Buyer Utility.}
We assume that each buyer's total utility is a discounted sum of their per-round utilities. Formally, let $\gamma_i \in (0,\bargam]$ denote the utility discount factor of buyer $i$, where $\bargam<1$ is a known constant. Then, buyer $i$’s total expected utility is defined as
\begin{align*}
    U_i(\B;\A, \E,\T) \coloneq \expec_{\omrgame} \sbra{\,\sum_{t \in \T_i} \gamma_i^{t-1} u(\theta_t, p_t, b_t)},\quad u(\theta, p, b) \coloneq (\theta - p) \cdot \ind[b \geq p].
    \numberthis\label{eq:buyerutil}
\end{align*}
Here $\expec_{\omrgame}$ denotes the expectation over all random variables realized during the execution of $\omrgame$, as described in Protocol~\ref{prot:omr}.

One may interpret the discount factor $\gamma_i$ as controlling the effective planning horizon of a buyer’s strategy, that is, how much the buyer values future surplus relative to immediate gains. A smaller $\gamma_i$ implies more myopic behavior, while values closer to $1$ reflect greater long-term planning, where the buyer is willing to incur short-term losses for greater future gains. Notably, in the extreme case of undiscounted utilities, where $\gamma_i = 1$, it is known that no algorithm can guarantee no-regret learning against a utility-maximizing buyer, even in the single-buyer, non-contextual setting~\citep{amin2013learning}. For this reason, we restrict our attention to the case $\gamma_i \in (0,\bargam]$ for some $\bargam < 1$. Requiring $\gamma_i>0$ ensures that a deviation in any round has positive weight in the discounted objective.

\paragraph{Nash Equilibrium.}
Given the definition of buyer utilities, we now formalize the notion of a Nash equilibrium in our setting, which captures a stable outcome where no buyer can unilaterally improve their utility. Let $\B_{-i} \coloneq (\B_j)_{j \in [n] \setminus \{i\}}$ denote the strategies of all buyers except buyer~$i$. We say that the strategy profile $\B \coloneq (\B_i)_{i \in [n]}$ is a \emph{Nash equilibrium} if
\begin{align*}
    \forall i \in [n], \quad U_i(\B_i, \B_{-i}; \A, \E,\T) = \sup_{\B_i}\,U_i(\B_i, \B_{-i}; \A, \E,\T).
\end{align*}
We denote by $\nash(\A, \E, \T, \gamma)$ the set of all Nash equilibria corresponding to a given seller algorithm $\A$, environment $\E$, round partition $\T$ and discount factors $\gamma \coloneq (\gamma_i)_{i \in [n]}$.

For our algorithms, which use a finite set of experts, $\nash$ is nonempty for every deterministic permitted environment. Quantized bids, the exploration prices of \S\ref{sec:multibuyer}, and the seller's other random choices have finite support, so the finite-horizon game has a finite game tree and hence a mixed Nash equilibrium. Since each quantized bid can be implemented by submitting that value itself, this equilibrium is also an equilibrium of the game with bids in $[0,1]$. Appendix~\ref{app:finitegame} gives the details.

\paragraph{Revenue and Regrets.}
The seller's total expected revenue is defined as
\begin{align*}
    \rev(\A;\B,\E,\T)
    &\coloneq\expec_{\omrgame}\sbra{\sum_{t=1}^T\irev(f_t,x_t,b_t)},\\
    \irev(f,x,b)&\coloneq f(x)\ind[b\geq f(x)].
    \numberthis\label{eq:sellerrev}
\end{align*}
Here $f_t=f_{w_t,a_t}$ is the policy selected before the current context is observed. Although prices may exceed $1$, the revenue in each round lies in $[0,1]$ because a sale requires $p_t\leq b_t\leq1$. We compare this revenue with the revenue of the best fixed policy in the same class, evaluated on the true values along the realized context--value path:
\begin{align*}
    \opt(\A;\B,\E,\T)\coloneq
    \expec_{\omrgame}\sbra{\sup_{f\in\F_{\textup{aff}}}
        \sum_{t=1}^T\irev(f,x_t,\theta_t)}.
\end{align*}

The difference between the above two quantities, commonly referred to as regret, depends on the buyers' strategies $\B$. We define two refined notions of regret, capturing performance guarantees under different behavioral assumptions on $\B$.

The first is the \emph{standard regret}, where all buyers report truthfully: $\widetilde b_t=\theta_t$, and hence $b_t=Q_h(\theta_t)$. Let $\Btru$ denote this truthful bidding strategy profile. Then, the standard regret is
\begin{align*}
    \reg(\A; \E,\T) \coloneq \opt(\A; \Btru, \E,\T) - \rev(\A; \Btru, \E,\T).
\numberthis\label{eq:reg}
\end{align*}
Our main focus is on the \emph{strategic regret}, which captures the worst-case efficiency loss due to strategic buyer behavior, as formalized by Nash equilibrium. It is defined as
\begin{align*}
    \sreg(\A; \E, \T, \gamma) \coloneq \sup_{\B \in \nash(\A, \E, \T, \gamma)} \Big[ \opt(\A; \B, \E,\T) - \rev(\A; \B, \E,\T) \Big].
\numberthis\label{eq:Sreg}
\end{align*}
In contrast to the standard regret~\eqref{eq:reg}, the above is a worst-case regret evaluated over \emph{all possible} Nash equilibria. As such, it closely aligns with the notion of \emph{Price-of-Anarchy}, as it quantifies the worst-case degradation in performance (revenue) due to selfish (strategic) behavior at equilibrium.

\paragraph{Hardness of OMR.}
Even with truthful bids, OMR under Protocol~\ref{prot:omr} faces a computational barrier. We prove the following result in Appendix~\ref{proof:OMRHardness} by reducing proper halfspace learning to affine pricing. The hardness of proper halfspace learning is due to~\citet{guruswami2009hardness}.
\vspace{0.5em}
\begin{restatable}{theoremm}{thmOMRHardness}
    Suppose $\mathrm{NP}\not\subseteq\mathrm{BPP}$, and let the bid precision $m(T)$ in Protocol~\ref{prot:omr} be polynomial in $T$, as in the choice~\eqref{eq:gridprecision} of our algorithms. No randomized seller algorithm can have both running time $\poly(d,T,\ell)$ and expected standard regret at most $C(d,\ell)T^z$, uniformly over environments and horizons, for a fixed $z<1$ and a polynomial $C$. Here $\ell$ bounds the bit length of each observation. This holds even for oblivious IID context--value pairs.
    \label{thm:OMRHardness}
\end{restatable}

As this complexity-theoretic barrier rules out computationally efficient algorithms with regret $C(d,\ell)T^z$, we instead aim to design algorithms that satisfy an approximate notion of no-regret. The polynomial-time approximation scheme of Definition~\ref{def:approx-no-regret} is the online analogue of the classical one, which finds a $(1-\eps)$-approximate solution in time polynomial in the input size for each fixed $\eps>0$. In our setting the approximation is additive: the regret may include a term $\eps T$, so the average revenue per round is within $\O(\eps)$ of that of the best fixed policy, up to a term that vanishes as $T$ grows. If both the running time and the coefficient of $T^z$ in the regret were polynomial in $1/\eps$, taking $\eps=T^{-c}$ for a small constant $c>0$ would give a polynomial-time algorithm with regret $\poly(d,\ell)\,T^{z'}$ for some $z'<1$. Thus, in the online setting, a scheme that is also polynomial in $1/\eps$ amounts to an efficient no-regret algorithm, which Theorem~\ref{thm:OMRHardness} rules out.
\vspace{0.5em}
\begin{definition}
    For a fixed $\eps>0$, a seller's algorithm $\A$ achieves \emph{$\eps$-approximate no-regret} if $\reg(\A; \E,\T) \in \O(T^z + \eps T)$ for some $z \in [0,1)$. The notion of \emph{$\eps$-approximate no-strategic-regret} is defined analogously by replacing $\reg$ with $\sreg$. Moreover, when an $\eps$-approximate no-regret algorithm runs in time polynomial in $T$ and $d$ for fixed $\eps$, we refer to it as a \emph{polynomial-time approximation scheme} (PTAS).
    \label{def:approx-no-regret}
\end{definition}

\subsection{Online Expert Algorithms}
Our algorithms treat any online expert algorithm as a black-box oracle.
For completeness, we briefly recall the standard definition of online expert learning and its associated regret (see~\citet{cesa2006prediction,hazan2016introduction,bubeck2015convex}).

The goal of an online expert algorithm is to compete with the best among $K$ experts, indexed by $z \in [K]$, when reward functions $r_t : [K] \rightarrow [0,1]$ are chosen adaptively by an environment based on past interactions. Formally, an (anytime) expert algorithm is a mapping $\L : \bigcup_{j=0}^{\infty} \big([0,1]^K\big)^j \rightarrow \Delta_{[K]}$, where the argument represents the sequence of past reward functions. At each round $t \in [T]$, the algorithm selects $z_t \sim \L(r_{\leq t-1})$, where $r_{\leq t-1} \coloneq (r_\tau)_{\tau \in [t-1]}$ denotes the history of rewards up to round $t-1$. The environment may also use current observations from outside the expert problem, but it fixes $r_t$ without observing $z_t$. We call such an environment non-anticipating.
Then, the regret of a $K$-expert algorithm $\L$ over $T$ rounds against environment $\E$ is defined as
\begin{align*}
    \reg^{\textup{exp}}(\L; \E) \coloneq \expec\sbra{\sup_{z \in [K]} \sum_{t=1}^T \big( r_t(z) - r_t(z_t) \big)},
    \numberthis\label{eq:ExpertRegret}
\end{align*}
which quantifies the expected difference between the cumulative reward of the best fixed expert in hindsight and that obtained by the algorithm.
Common examples of online expert algorithms include Hedge~\citep{Freund1997119} and Follow-the-Perturbed-Leader (FTPL)~\citep{kalai2005efficient, hutter2005adaptive}.

\section{Online Myersonian Regression with Truthful Buyers}\label{sec:truthful}
To build toward strategy-robust learning, we begin with truthful bidding, $\widetilde b_t=\theta_t$, so that the quantized bids are $b_t=Q_h(\theta_t)$. Our algorithm $\A^\L_\omr$ reduces OMR to online expert learning, translating the expert regret guarantee~\eqref{eq:ExpertRegret} into a standard regret guarantee~\eqref{eq:reg}. We extend this construction to strategic buyers in \S\ref{sec:multibuyer}.

\paragraph{Key Technical Challenge.}
The reduction must remain valid when the observed bids depend on past prices. With truthful bids, this happens when a permitted environment adapts its pairs to past public prices. With strategic bids (\S\ref{sec:multibuyer}), it happens even when the true context--value sequence is oblivious. Fixed dimensionality reductions, such as Johnson--Lindenstrauss sketching~\citep{johnson1984extensions}, work for nonadaptive data, but their usual guarantees do not cover adaptive inputs~\citep{hardt2013how,cohen2022robustness}. We instead construct experts from past contexts, with an approximation guarantee that holds along every realized path.

\paragraph{Algorithm Overview.}
We begin by describing the algorithm, and then prove our main result, Theorem~\ref{thm:RegretOMR}.
Our algorithm uses oracle access to an online expert algorithm $\L$ over a finite expert set $\Z_{\eps,T}$, defined below. Each expert specifies a few coefficients on past contexts and a bias. Together with the past contexts, these determine the expert's pricing policy for round $t$ before $x_t$ is observed.

For $\eps\in(0,1/2]$, define
\begin{align*}
    M_\eps&\coloneq\ceil{256\eps^{-2}},\qquad
    N_\eps\coloneq\ceil{24/\eps},\\
    J_\eps&\coloneq\cbra{-1+3j/N_\eps:j=0,\dots,N_\eps},\\
    \Z_{\eps,T}&\coloneq\bigg\{
        \big((\beta_\tau)_{\tau\in S},a\big):
        S\subseteq[T],\ |S|\leq M_\eps,\ a\in J_\eps,\\
        &\hspace{0.65in}
        \forall\tau\in S,\quad
        \beta_\tau\in[-8/\eps,8/\eps]\cap(\eps/32)\zz
    \bigg\}.
    \numberthis\label{eq:Zset}
\end{align*}
Write $a(z)$ for the bias component of $z$. Enumerating index sets, coefficients, and bias values gives
$|\Z_{\eps,T}|\leq(T/\eps)^{\O(\eps^{-2})}$. This is polynomial in $T$ for fixed $\eps$, in contrast to the exponential dependence on the weight dimension $d$ that arises when directly searching over a discretization of $\bb^d$.
For each $z=((\beta_\tau)_{\tau\in S},a)\in\Z_{\eps,T}$, set
\begin{align*}
    v_t(z;x_{<t})\coloneq
    \frac{\sum_{\tau\in S\cap[t-1]}\beta_\tau x_\tau}
    {\max\cbra{1,\left\|\sum_{\tau\in S\cap[t-1]}\beta_\tau x_\tau\right\|}}
    \in\bb^d.
    \numberthis\label{eq:vtfunc}
\end{align*}
The expert's pricing policy in round $t$ is
\begin{align*}
    f_t(z;x)\coloneq
    [\innp{v_t(z;x_{<t}),x}+a(z)-\eps/2]_+.
    \numberthis\label{eq:expertpolicy}
\end{align*}
Lowering the bias by $\eps/2$ preserves membership in the affine class, so $f_t(z;\cdot)\in\F_{\textup{aff}}$. For fixed $z$, the weight vector is constructed using only contexts through $t-1$. After observing the quantized bid, the seller feeds the reward vector $r_t(z)=\irev(f_t(z;\cdot),x_t,b_t)$ to $\L$. Algorithm~\ref{alg:omr} and the strategic algorithm in \S\ref{sec:multibuyer} both set the bid precision to
\begin{align*}
    m\coloneq\ceil{\max\cbra{T^2,1024\eps^{-3}}},\qquad h\coloneq1/m.
    \numberthis\label{eq:gridprecision}
\end{align*}

\vspace{0.5em}
\begin{restatable}{theoremm}{thmRegretOMR}
    Let $\L$ have expected regret at most $R(T\log K)$ over $K$ experts against non-anticipating adaptive environments. For $\eps\in(0,1/2]$, any horizon $T$, and any permitted environment $\E$,
    \begin{align*}
        \reg(\A^\L_\omr;\E,\T)
        \leq R(T\log|\Z_{\eps,T}|)
        +(3\eps/4)T+\min\{T,M_\eps\}+hT.
    \end{align*}
    In particular, with Hedge,
    \begin{align*}
        \reg(\A^\L_\omr;\E,\T)
        \in\O\big(\eps^{-1}\sqrt{T\log(T/\eps)}+\eps T\big).
    \end{align*}
    The running time is $(T/\eps)^{\O(\eps^{-2})}\poly(d,T,1/\eps)$, giving a PTAS for OMR with truthful bids.
    \label{thm:RegretOMR}
\end{restatable}

\begin{algorithm2e}[t]
    \caption{$\A^\L_\omr$: Online Myersonian Regression Algorithm for Truthful Buyers}\label{alg:omr}
    \SetKwComment{Comment}{\hfill$\triangleright$~}{}
    \DontPrintSemicolon
    \textbf{Oracle:} Online Expert Algorithm $\L$.\\
    \textbf{Input:} Horizon $T$; approximation parameter $\eps>0$.\\
    Set $\Z_{\eps,T}$ as the expert set of $\L$ and choose $h$ by~\eqref{eq:gridprecision}.\\
    \For{$t\in[T]$}{
        Sample $z_t\sim\L(r_{\leq t-1})$ and privately select
        $f_t\gets f_t(z_t;\cdot)$ before observing $x_t$.\\
        Observe $x_t$ and receive the sealed bid $\widetilde b_t$.\\
        Set $b_t\gets Q_h(\widetilde b_t)$, reveal $p_t\gets f_t(x_t)$, and sell if $p_t\leq b_t$.\\
        Feed $r_t(z)\gets\irev(f_t(z;\cdot),x_t,b_t)$, for $z\in\Z_{\eps,T}$, to $\L$.\\
        \Comment*{{\rmfamily The expert functions are defined in~\eqref{eq:expertpolicy}.}}
    }
\end{algorithm2e}

\paragraph{Proof of Theorem~\ref{thm:RegretOMR}.}
The core ingredient is an \emph{online sketch} of a fixed weight vector $w\in\bb^d$ by an expert $z$ whose weight vector $v_t(z;x_{<t})$ is constructed from past contexts. The approximation may fail in a bounded number of rounds, whose cost we charge separately.
\begin{restatable}{lemmma}{lemOnlineSketch}\emph{[Online Sketch]}
    For any context sequence $(x_t)_{t\in[T]}$ in $\bb^d$ and any $w\in\bb^d$, there is an expert $z\in\Z_{\eps,T}$ as defined in~\eqref{eq:Zset} and a set $D\subseteq[T]$, with $|D|\leq\min\{T,M_\eps\}$, such that
    \begin{align*}
        \forall t\notin D,\qquad
        \big|\innp{v_t(z;x_{<t}),x_t}-\innp{w,x_t}\big|
        \leq\eps/8.
    \end{align*}
    The bias component of $z$ may be any element of $J_\eps$.
    \label{lem:OnlineSketch}
\end{restatable}

Combining the online sketch with a grid of bias values and lowering the bias by $\eps/2$, as in~\eqref{eq:expertpolicy}, gives the following revenue approximation bound for the finite expert set, proved in Appendix~\ref{proof:Zapprox}. Outside the set $D$ of Lemma~\ref{lem:OnlineSketch}, the shift keeps the expert's price at or below the price of the affine policy that it approximates, so the expert sells whenever that policy does.
\vspace{0.5em}
\begin{restatable}{lemmma}{lemZapprox}
    For any $(x_t,b_t)_{t\in[T]}$ with $x_t\in\bb^d$, $b_t\in[0,1]$, and $\eps\in(0,1/2]$,
    \begin{align*}
        \sup_{f\in\F_{\textup{aff}}}\sum_{t=1}^T\irev(f,x_t,b_t)
        \leq
        \sup_{z\in\Z_{\eps,T}}\sum_{t=1}^T
            \irev(f_t(z;\cdot),x_t,b_t)
        +(3\eps/4)T+\min\{T,M_\eps\}.
    \end{align*}
    \label{lem:Zapprox}
\end{restatable}

Truthful bidding gives $b_t\geq\theta_t-h$, and lowering a comparator's bias by $h$ yields a policy whose revenue on the quantized bids is at least the comparator's revenue on the true values minus $hT$. Lemma~\ref{lem:RevStab} formalizes this argument. Applying Lemma~\ref{lem:Zapprox} and the oracle's regret bound then gives
\begin{align*}
    \reg(\A^\L_\omr;\E,\T)
    &\leq
    \expec_{\omrgameTruthful}\sbra{
       \sup_{z\in\Z_{\eps,T}}\sum_{t=1}^T(r_t(z)-r_t(z_t))}
       \\
    &\quad +(3\eps/4)T+\min\{T,M_\eps\}+hT\\
    &\leq R(T\log|\Z_{\eps,T}|)
       +(3\eps/4)T+\min\{T,M_\eps\}+hT.
\end{align*}
For Hedge, use $\log|\Z_{\eps,T}|=\O(\eps^{-2}\log(T/\eps))$,
$\min\{T,M_\eps\}\leq\sqrt{TM_\eps}$, and $hT\leq1/T$.
Enumerating the experts and computing their rewards gives the stated running time.
$\blacksquare$

\paragraph{IID Online-to-Batch Conversion.}
If the pairs $(x_t,\theta_t)$ are IID with distribution $P$, choose $J\sim U_{[T]}$ once after training and return the policy $f_J$ selected in round $J$. This policy belongs to $\F_{\textup{aff}}$ in every realization. Let $\rev_P(f)$ denote the expected revenue of a policy $f$ when $(X,\Theta)\sim P$. Since each policy is selected before observing the current context and value, we obtain
\begin{align*}
    \expec\sbra{\sup_{f\in\F_{\textup{aff}}}\rev_P(f)-\rev_P(f_J)}
    \leq \reg(\A^\L_\omr;\E,\T)/T.
\end{align*}
Here the expectation is over the training run and $J$. This conversion applies to truthful bids. Appendix~\ref{app:onlinetobatch} gives the short proof.

\paragraph{Proof Sketch of Lemma~\ref{lem:OnlineSketch}.}
We conclude this section with a proof sketch of Lemma~\ref{lem:OnlineSketch} and defer the full proof to Appendix~\ref{proof:OnlineSketch}.
Fix a comparator $w\in\bb^d$. We construct its sketch explicitly; the construction is only an existence argument and does not require the seller to know $w$. Initialize $u=v=0$. At the start of round $t$, record $v_t\gets v$. Then, while $|\innp{v,x_t}-\innp{w,x_t}|>\eps/8$, update
\begin{align*}
    u\gets u-(\eps/32)\sign(\innp{v,x_t}-\innp{w,x_t})x_t,
    \qquad v\gets\frac{u}{\max\{1,\|u\|\}}.
    \numberthis\label{eq:OSupdate-main-text}
\end{align*}
The recorded $v_t$ is not changed by these updates, so each update affects predictions only from round $t+1$ onward.

Index the updates by $i$, let $t_i$ be the round containing update $i$, and let $v_i^{\textup{upd}}$ be the vector immediately before that update. The update sequence is lazy OGD on the convex losses
$\ell_i(v)=|\innp{v,x_{t_i}}-\innp{w,x_{t_i}}|$, with step size $\eps/32$.
Each update incurs loss greater than $\eps/8$, while $\ell_i(w)=0$.
Lemma~\ref{lem:LazyOGD} gives, for any number $M\geq1$ of updates,
\begin{align*}
    M\eps/8<\sum_{i=1}^M\ell_i(v_i^{\textup{upd}})
    \leq16/\eps+(\eps/16)M,
\end{align*}
so $M<256\eps^{-2}$. This also proves termination of the update loop.
Hence there are at most $M$ exceptional rounds. Aggregating the updates made on the same context produces coefficients in $(\eps/32)\zz$ of magnitude at most $8/\eps$, and unrolling the updates made before the current round gives exactly~\eqref{eq:vtfunc}. $\blacksquare$

\section{Strategy-robust Regret Guarantee}\label{sec:multibuyer}
In this section, we introduce a simple modification to $\A^\L_\omr$, called the \textit{Sparse Update Mechanism (SUM)}, which makes the algorithm strategy-robust. When the oracle $\L$ is no-regret, the resulting algorithm, given in Alg.~\ref{alg:sum-omr}, achieves $\eps$-approximate no-strategic-regret (Def.~\ref{def:approx-no-regret}).
We retain the expert policies from~\eqref{eq:expertpolicy}, draw exploration prices uniformly from $\G_h$, and update $\L$ only on a private random subset of rounds. We choose $h$ by~\eqref{eq:gridprecision} and set
\begin{align*}
    \rho\coloneq
    \min\cbra{1,\frac{(1-\bargam)\eps^3}{2048\bargam}}.
    \numberthis\label{eq:updateprob}
\end{align*}

\begin{algorithm2e}[t]
\caption{$\A^\L_\sm$: Strategy-robust Online Myersonian Regression via Sparse Update}\label{alg:sum-omr}
\SetKwComment{Comment}{\hfill$\triangleright$~}{}
\DontPrintSemicolon
\textbf{Oracle:} Online Expert Algorithm $\L$.\\
\textbf{Input:} Horizon $T$; approximation parameter $\eps>0$; discount bound $\bargam$.\\
Set $\Z_{\eps,T}$ as the expert set of $\L$; choose $h$ and $\rho$ by~\eqref{eq:gridprecision} and~\eqref{eq:updateprob}.\\
\For{$t\in[T]$}{
    Privately sample $\omega_t\sim\ber(\eps)$ before observing $x_t$.\\
    \If{$\omega_t=1$}{
        Sample $\lam\sim U_{\G_h}$ and select $(w_t,a_t)\gets(0,\lam)$.\\
        \Comment*{{\rmfamily A constant pricing policy, chosen before the context.}}
    }
    \Else{
        Sample $z_t\sim\L((r_\tau)_{\tau<t:\xi_\tau=1})$ and select
        $(w_t,a_t)\gets(v_t(z_t;x_{<t}),a(z_t)-\eps/2)$.
    }
    Observe $x_t$ and receive the sealed bid $\widetilde b_t$.\\
    Set $b_t\gets Q_h(\widetilde b_t)$; reveal $p_t\gets[\innp{w_t,x_t}+a_t]_+$ and sell if $p_t\leq b_t$.\\
    Set $r_t(z)\gets\irev(f_t(z;\cdot),x_t,b_t)$ for each $z\in\Z_{\eps,T}$.\\
    Privately sample a fresh $\xi_t\sim\ber(\rho)$, independently of the interaction so far.\\
    \If{$\xi_t=1$}{Feed $r_t$ to $\L$.}
}
\end{algorithm2e}

\paragraph{Algorithm Overview.}
The key idea behind SUM is to limit buyer influence by randomly selecting—privately and independently—which rounds are used to update the seller’s learning algorithm. Specifically, in each round, the seller flips a biased private coin $\xi_t$, the update coin, to decide whether to feed the observed interaction back to the expert algorithm $\L$. As a result, only a sparse, random subset of rounds contributes to learning, making it difficult for any buyer to reliably influence future prices.

In addition, the seller occasionally selects a price uniformly at random: with probability $\eps$, it uses the constant policy $f(x)=\lambda$ with $\lambda\sim U_{\G_h}$. Like every exploitation policy, this policy is chosen before the current context is observed, and both the policy and the decision to explore remain hidden until the bid is submitted. As shown in Lemma~\ref{lem:RandomPricing}, random pricing penalizes bids far from the true value, while sparse updating limits the benefit of manipulating future prices. Together they make every Nash equilibrium approximately truthful (Lemma~\ref{lem:SparseUpdateNash}).

This approach of combining randomized updates and random pricing is conceptually simpler than prior work based on differential privacy~\citep{liu2018learning,abernethy2019learning}, and it supports a black-box reduction from any no-regret online expert algorithm.
The regret guarantee for Alg.~\ref{alg:sum-omr} is given below:

\vspace{0.5em}
\begin{restatable}{theoremm}{thmSRegretOMR}
    Let $\L$ have expected regret at most $R(n\log K)$ after every $n$ rounds against non-anticipating adaptive environments, where $R:[0,\infty)\to[0,\infty)$ is nonnegative, nondecreasing, and concave. For $\eps\in(0,1/4]$, $T\geq1$, $0<\gamma_i\leq\bargam<1$, and any permitted environment $\E$,
    \begin{align*}
        \sreg(\A^\L_\sm;\E,\T,\gamma)
        &\leq \rho^{-1}R(\rho T\log|\Z_{\eps,T}|)\\
        &\quad+\O\big(\sqrt{\rho^{-1}T\log(|\Z_{\eps,T}|T)}+\eps T\big)
        +\min\{T,M_\eps\},
    \end{align*}
    with $\rho$ from~\eqref{eq:updateprob}. In particular, with Hedge,
    \begin{align*}
        \sreg(\A^\L_\sm;\E,\T,\gamma)
        \in\O\bigg(
          \bigg(\eps^{-1}
          +\eps^{-5/2}\sqrt{\frac{\bargam}{1-\bargam}}\bigg)
          \sqrt{T\log(T/\eps)}+\eps T
        \bigg).
    \end{align*}
    The running time is $(T/\eps)^{\O(\eps^{-2})}\poly(d,T,1/\eps)$, giving a PTAS for OMR with strategic buyers.
    \label{thm:SRegretOMR}
\end{restatable}

\paragraph{Proof Sketch of Theorem~\ref{thm:SRegretOMR} (Full Proof in Appendix~\ref{app:thm:SregretOMR}).}
We first bound a buyer's potential gain in future utility from changing its bid. Fix a buyer $i$ and a round $t^\star\in\T_i$, and consider a modified strategy $\B_i'$ that submits a different bid at $t^\star$ but records, in place of that bid, the bid that $\B_i$ originally sampled. Once the price is revealed, it also records the allocation and utility that the original bid, after quantization, would have produced. In later rounds, it runs $\B_i$ on this record and appends the actual new observations together with the bids and private random variables that $\B_i$ generates. We call the resulting record the buyer's virtual private history, denoted by $\widehat{\mathcal H}_{i,t}$ after round $t$. The buyer can construct it from its original bid, the revealed price, its value, and subsequent observations.

The effect of the changed bid depends on the update coin $\xi_{t^\star}$. If $\xi_{t^\star}=0$, the changed bid is not fed to $\L$, affects no other part of the seller's pricing state, and is not observed by the other buyers or by $\E$. Running $\B_i$ on the virtual private history $\widehat{\mathcal H}_{i,t}$ then reproduces the original continuation, and the other parties see the same public path. If $\xi_{t^\star}=1$, the buyer's future discounted utility can change by at most $2\bargam/(1-\bargam)$ times the current discount weight $\gamma_i^{t^\star-1}$. Since $\xi_{t^\star}=1$ has probability $\rho$, the expected future benefit is at most $2\rho\bargam/(1-\bargam)$ times this weight.

Under random pricing on $\G_h$, bidding sufficiently far from the true value incurs an expected utility loss in the current round. The following lemma, proved in Appendix~\ref{proof:RandomPricing}, quantifies this loss. It also shows that, at any fixed nonnegative price $p$, the quantized truthful bid $Q_h(\theta)$ achieves utility within $h$ of the maximum $u(\theta,p,\theta)$.
\vspace{0.5em}
\begin{restatable}{lemmma}{lemRandomPricing}
    For $\theta\in[0,1]$, $b\in\G_h$, and $\lambda\sim U_{\G_h}$,
    \begin{align*}
        \expec_{\lambda}\sbra{
            u(\theta,\lambda,Q_h(\theta))-u(\theta,\lambda,b)}
        \geq
        \frac{[|\theta-b|-2h]_+^2}{2(1+h)}.
    \end{align*}
    At any nonnegative price $p$, using the quantized truthful bid $Q_h(\theta)$ loses at most $h$ in utility relative to $u(\theta,p,\theta)$.
    \label{lem:RandomPricing}
\end{restatable}

Set
\begin{align*}
    \del_h\coloneq
    2h+\sqrt{\frac{8(h+2\rho\bargam/(1-\bargam))}{\eps}}.
    \numberthis\label{eq:neartruthradius}
\end{align*}
For $|\theta-b|>\del_h$, the expected gain in current utility from bidding truthfully exceeds the expected future benefit from misreporting. Hence, if such a bid occurred with positive probability, switching to the truthful bid on that event would be a profitable deviation, contradicting Nash equilibrium.
\vspace{0.5em}
\begin{restatable}{lemmma}{lemSparseUpdateNash}\emph{[Sparse Update Mechanism]}
    For any $\B\in\nash(\A^\L_\sm,\E,\T,\gamma)$ with $0<\gamma_i\leq\bargam<1$,
    \begin{align*}
        \Pr_{\omrgameSR}\sbra{
            \forall t\in[T],\ |\theta_t-b_t|\leq\del_h}=1.
    \end{align*}
    The choices~\eqref{eq:gridprecision} and~\eqref{eq:updateprob} give $\del_h<\eps/4$.
    \label{lem:SparseUpdateNash}
\end{restatable}

By Lemma~\ref{lem:SparseUpdateNash}, proved in Appendix~\ref{proof:SparseUpdateNash}, every quantized bid lies within $\del_h$ of the true value almost surely. Consequently, the benchmark revenue computed from these bids falls short of the benchmark revenue on true values by at most $\del_h$ per round, as the following lemma shows. Its proof uses the fact that lowering the bias preserves membership in the policy class: $[f_{w,a}(x)-\del]_+=f_{w,a-\del}(x)$.
\vspace{0.5em}
\begin{restatable}{lemmma}{lemRevStab}
    For any $(x_t,\theta_t,b_t)_{t\in[T]}$ with $\theta_t,b_t\in[0,1]$ and $b_t\geq\theta_t-\del$, where $\del\geq0$,
    \begin{align*}
        \sup_{f\in\F_{\textup{aff}}}\sum_{t=1}^T\irev(f,x_t,\theta_t)
        \leq
        \sup_{f\in\F_{\textup{aff}}}\sum_{t=1}^T\irev(f,x_t,b_t)
        +\del T.
    \end{align*}
    \label{lem:RevStab}
\end{restatable}

Applying Lemma~\ref{lem:RevStab} with $\del=\del_h$ transfers the benchmark from the true values to the quantized bids, and Lemma~\ref{lem:Zapprox} then transfers it to the finite expert set, accounting for the loss in the exceptional rounds. Finally, Lemma~\ref{lem:SparseUpdateRegret} in Appendix~\ref{app:thm:SregretOMR} bounds the regret under sparse feedback, and exploration costs at most $\eps T$ in expectation. Combining these bounds yields Theorem~\ref{thm:SRegretOMR}. $\blacksquare$

\section{Conclusion}\label{sec:conclusion}
In this work, we studied online contextual pricing with multiple strategic buyers against a benchmark of fixed affine pricing policies. We introduced a notion of strategy-robust regret that captures performance at the worst Nash equilibrium, in the spirit of the Price-of-Anarchy. We proposed Online Myersonian Regression (OMR) and showed that achieving no-regret learning in this model is computationally hard unless $\mathrm{NP}\subseteq\mathrm{BPP}$, even when buyers are truthful. To address this barrier, we developed a PTAS for truthful buyers, based on an online sketch constructed from past contexts, with running time $(T/\eps)^{\O(\eps^{-2})}\poly(d,T,1/\eps)$.

To extend robustness to multi-buyer strategic settings, we introduced the \emph{Sparse Update Mechanism (SUM)}—a simple randomization technique in which the seller updates its pricing policy using feedback from only a sparse subset of rounds and occasionally posts a random price. This effectively limits buyer influence and ensures regret guarantees that hold uniformly over \emph{all Nash equilibria}. Our approach is conceptually simple and broadly applicable, as it is based on a black-box reduction from any no-regret expert algorithm. We believe this framework can be further extended to yield no-regret guarantees under more refined benchmarks, such as dynamic or adaptive regret~\citep{hazan2016introduction, cesa2006prediction}.

\section*{Acknowledgments}
This work was partially supported by the National Science Foundation (NSF)
under Grant IIS-2441796.

\bibliographystyle{plainnat}
\bibliography{references}

\appendix
\section{Proof of Hardness for No-Regret Learning in OMR}\label{app:hardness}
We prove hardness for the same class of affine pricing policies used by the seller and the benchmark, by a reduction from proper halfspace learning. All simulated observations have polynomial bit length, so $\ell$ in Theorem~\ref{thm:OMRHardness} is polynomial in the instance size.

\begin{propositionn}[Affine Pricing Gap]\label{thm:AffineGap}
There is a polynomial-time reduction that maps each instance of an NP-hard promise problem to an efficiently samplable distribution $P$ of rational pairs $(X,\Theta)$, with $\|X\|=1$ and $\Theta\in[0,1]$, together with rational numbers $c_0=1/8$ and $\kappa>0$, where $\kappa^{-1}$ is polynomial in the instance size, such that:
\begin{align*}
    \textup{YES:}\quad&
       \sup_{f\in\F_{\textup{aff}}}
           \expec_P[\irev(f,X,\Theta)]\geq c_0+11\kappa/16,\\
    \textup{NO:}\quad&
       \sup_{f\in\F_{\textup{aff}}}
           \expec_P[\irev(f,X,\Theta)]\leq c_0+5\kappa/16.
\end{align*}
\end{propositionn}

\begin{proofof}{Proposition~\ref{thm:AffineGap}.}
\paragraph{Halfspace Gap and Normalization.}
By Theorem 8.1 of~\citet{guruswami2009hardness}, it is NP-hard to distinguish multisets of labeled Boolean examples $(z,s)\in\{-1,1\}^D\times\{-1,1\}$ for which some halfspace has agreement at least $7/8$ from those for which every affine halfspace has agreement at most $5/8$. We use the completeness assignment of their reduction, a zero-one vector $A$ with threshold zero. In their paper, Section 4 constructs this assignment and Section 6 copies its coordinates, and on the satisfied tuples the two checks in their Section 7 have strictly positive and strictly negative integer scores. Hence $s\innp{A,z}\geq1$ on at least $7/8$ of the output examples; this integer margin is the additional property we need.

Let $k=\ceil{\sqrt D}$, pad $z$ to dimension $k^2$ using coordinates equal to $1$, and pad $A$ with zeros. Put
\begin{align*}
    u=z_{\textup{pad}}/k,\qquad
    v=A_{\textup{pad}}/k,\qquad L=k^2.
\end{align*}
Then $\|u\|=1$, $\|v\|\leq1$, and $L\innp{v,u}=\innp{A,z}$.
Every affine halfspace on $u$ induces one on $z$, so the NO guarantee is preserved.

\paragraph{A Revenue Representation.}
Define
\begin{align*}
    c_0\coloneq1/8,\qquad
    \kappa\coloneq1/(256L),\qquad
    q_L(p)\coloneq
       \max\{-1,\min\{1,16L(p-1/2)\}\}.
\end{align*}
For a source label $s$, first define an ideal conditional distribution of $\Theta$ by its survival function at positive prices:
\begin{align*}
    S_s(p)\coloneq\Pr(\Theta\geq p\mid s)=
    \begin{cases}
        4(c_0-\kappa s),&0<p<1/4,\\
        (c_0+\kappa s q_L(p))/p,&1/4\leq p\leq1.
    \end{cases}
    \numberthis\label{eq:hardnesssurvival}
\end{align*}
There is an atom $1-S_s(0+)$ at zero and an atom $S_s(1)$ at one.
To verify that this defines a distribution, note that outside the ramp of $q_L$ the numerator of the second branch is a positive constant, while inside the ramp
\begin{align*}
    S_s(p)=16L\kappa s+\frac{c_0-8L\kappa s}{p},
\end{align*}
where $c_0-8L\kappa s\geq3/32>0$. Thus the survival function is nonincreasing, its pieces agree at their boundaries, and
$S_s(0+)=1/2-4\kappa s\in(0,1)$, $S_s(1)=1/8+\kappa s\in(0,1)$.
The ideal expected revenue $R_s(p)$ at price $p$, which equals $pS_s(p)$ for $p>0$, satisfies
\begin{align*}
    R_s(p)&=c_0+\kappa s q_L(p),&&1/4\leq p\leq1,\\
    R_s(p)&\leq c_0+\kappa s q_L(p),&&p\geq0.
    \numberthis\label{eq:hardnessbridge}
\end{align*}
For $0<p<1/4$, this follows from $q_L(p)=-1$ and
$R_s(p)=4p(c_0-\kappa s)$. At $p=0$, revenue is zero.
At $p>1$, actual revenue is zero while the right side is positive.

Under $P$, the source example $(z,s)$ is uniform over the multiset, and the context is the rational unit vector
\begin{align*}
    x(u)\coloneq(3u/5,4/5).
\end{align*}
Given $s$, the value $\Theta$ follows the conditional distribution above. A finite-precision version is given below.

\paragraph{Completeness.}
In the YES case, consider the policy with zero bias and weight vector
\begin{align*}
    w_v\coloneq(5v/24,5/8),\qquad
    \|w_v\|^2\leq125/288<1.
\end{align*}
Its price is
\begin{align*}
    p_v(x(u))=1/2+\innp{v,u}/8\in[3/8,5/8].
\end{align*}
Thus equality holds in~\eqref{eq:hardnessbridge}. Moreover,
\begin{align*}
    \frac{1+q_L(p_v(x(u)))}2
    =\sbra{\min\{1,1/2+L\innp{v,u}\}}_+.
\end{align*}
On the source examples with $s\innp{A,z}\geq1$, which comprise at least $7/8$ of the multiset, this gives $q_L(p_v(x(u)))=s$; hence $\expec[s\,q_L(p_v(X))]\geq7/8-1/8=3/4$ and the ideal expected revenue is at least $c_0+3\kappa/4$.

\paragraph{Soundness.}
Fix any $f\in\F_{\textup{aff}}$, and put
$g_f(u)\coloneq(1+q_L(f(x(u))))/2$.
For every threshold $\tau\in(0,1)$, the classifier
$\ind[g_f(u)\geq\tau]$ is an affine halfspace on the original $z$:
the corresponding price threshold lies strictly between $0$ and $1$, so lower clipping does not alter its defining affine inequality.
By averaging these threshold classifiers,
\begin{align*}
    \expec\bigg[
       \frac{1+s}{2}g_f(u)+\frac{1-s}{2}(1-g_f(u))
    \bigg]\leq5/8.
\end{align*}
Using~\eqref{eq:hardnessbridge}, the expected revenue under the ideal value distribution is at most
$c_0+\kappa(2(5/8)-1)=c_0+\kappa/4$.
This includes prices above $1$. The same bound holds when $f$ is drawn independently of $(X,\Theta)$ from any distribution over $\F_{\textup{aff}}$.

\paragraph{Finite Precision.}
We approximate the ideal conditional distribution of $\Theta$ using finitely many random bits.
Let $p_\pm=1/2\pm1/(16L)$ and sample $U$ uniformly from
$\{2^{-B},2\cdot2^{-B},\dots,1\}$.
If $U>S_s(0+)$, output $0$; if $U\leq S_s(1)$, output $1$.
Otherwise output
\begin{align*}
    \Theta=
    \begin{cases}
        (c_0-\kappa s)/U,&U\geq S_s(p_-),\\[1mm]
        (c_0-8L\kappa s)/(U-16L\kappa s),
           &S_s(p_+)\leq U<S_s(p_-),\\[1mm]
        (c_0+\kappa s)/U,&S_s(1)<U<S_s(p_+).
    \end{cases}
\end{align*}
These rational branches agree at their boundaries. Monotonic inversion gives a survival probability within $2^{-B}$ of~\eqref{eq:hardnesssurvival} at every positive price. Hence the expected revenue error is at most $\nu=2^{-B}$ uniformly over all nonnegative prices, including prices at which no sale occurs.
Choose $B$ so $\nu\leq\kappa/16$.
The YES bound becomes $c_0+11\kappa/16$ and the NO bound becomes $c_0+5\kappa/16$.
Sampling takes polynomial time, and each observation has bit length polynomial in the instance size. This proves the proposition.
\end{proofof}

\vspace{1em}
\thmOMRHardness*
\begin{proofof}{Theorem~\ref{thm:OMRHardness}.}\label{proof:OMRHardness}
Suppose such a seller algorithm exists. Given an instance of the promise problem, apply Proposition~\ref{thm:AffineGap} and run the algorithm for $T$ rounds on IID samples $(X_t,\Theta_t)$ from $P$, simulating a truthful buyer by supplying the quantized bid $Q_h(\Theta_t)$. In each round, the seller's policy $f_t\in\F_{\textup{aff}}$ is selected before the fresh pair is observed. Since $Q_h(\Theta_t)\leq\Theta_t$, the revenue of $f_t$ evaluated on the true value is at least its revenue on the quantized bid, pointwise. The claimed regret guarantee therefore implies, in the YES case,
\begin{align*}
    \expec\sbra{\frac1T\sum_{t=1}^T\irev(f_t,X_t,\Theta_t)}
    \geq c_0+11\kappa/16-C(d,\ell)T^{z-1}.
\end{align*}
Here we used that the expected empirical optimum is at least $T$ times the fixed YES comparator's expected revenue under $P$.
Choose polynomially large $T$ so that $C(d,\ell)T^{z-1}\leq\kappa/16$.
The resulting YES expected average is at least $c_0+5\kappa/8$.

In the NO case, conditional on the past, each current policy is independent of the fresh pair, so Proposition~\ref{thm:AffineGap} bounds its expected revenue evaluated on true values by $c_0+5\kappa/16$, including when the policy is selected using private randomness. The same upper bound therefore holds for the expected average revenue of the whole run.

Each run's average revenue evaluated on true values lies in $[0,1]$. Repeating independently $\O(\kappa^{-2})$ times estimates its expectation within $\kappa/16$ with probability at least $2/3$, by Hoeffding's inequality. Comparing with $c_0+\kappa/2$ distinguishes the source YES and NO instances in randomized polynomial time. This would imply $\mathrm{NP}\subseteq\mathrm{BPP}$.

The choice of $T$ is polynomial because $1-z>0$ is fixed, $C$ is polynomial, and $\kappa^{-1}$ is polynomial in the instance size. Since the bid precision $m(T)$ is polynomial in $T$, it is polynomial in the instance size as well. The contradiction proves the theorem.
\end{proofof}

\section{Proof of Key Lemmas}\label{app:keylemmas}
In this section, we prove two key lemmas: Lemma~\ref{lem:OnlineSketch}, which analyzes our online sketching technique used to establish the PTAS for OMR with truthful buyers, and Lemma~\ref{lem:SparseUpdateNash}, which shows the approximate truthfulness of \emph{any Nash equilibrium} induced by $\A^\L_\sm$ (Alg.~\ref{alg:sum-omr}).
\subsection{Proof of Lemma~\ref{lem:OnlineSketch}}
\paragraph{Preliminary.}
Before proving Lemma~\ref{lem:OnlineSketch}, we state a technical result that we use: a regret bound for Lazy Online Gradient Descent (Lazy OGD), given in Lemma~\ref{lem:LazyOGD}. This bound is a special case of the regret guarantee for Lazy Online Mirror Descent when applied with the Bregman divergence $\frac{1}{2}\|x - y\|^2$. For completeness, we provide a direct proof in Appendix~\ref{app:lazyOGD}.
\vspace{1em}
\begin{restatable}{lemmma}{lemLazyOGD}
    Let $M\in\nat$, and for each $i \in [M]$, let $f_i : \bb^d \rightarrow \rr$ be any convex loss function with unit-bounded subgradients, i.e., $\|\nabla f_i\| \leq 1$. Then, for any $\beta > 0$ and $v^\star \in \bb^d$, \begin{align*}
        \sum_{i \in [M]} \big(f_i(v_i) - f_i(v^\star)\big) \leq (2\beta)^{-1} + 2\beta M,
        \numberthis\label{eq:LazyOGDRegret}
    \end{align*} where $v_i$ is defined recursively as
    \begin{align*}
        u_{i+1} \coloneq u_i - \beta\,\nabla f_i(v_i), \quad u_1 = 0\quad \textup{and}\quad v_i \coloneq \frac{u_i}{\max\{1,\|u_i\|\}}.
        \numberthis\label{eq:LazyOGDUpdateRule}
    \end{align*}
    \label{lem:LazyOGD}
\end{restatable}
With this ingredient, we now prove Lemma~\ref{lem:OnlineSketch}, restated below for convenience.
\vspace{1em}
\lemOnlineSketch*
\begin{proofof}{Lemma~\ref{lem:OnlineSketch}.}\label{proof:OnlineSketch}
Fix $(x_t)_{t\in[T]}$ and $w\in\bb^d$. Define
\begin{align*}
    h_t(v)\coloneq\innp{v,x_t}-\innp{w,x_t}.
\end{align*}
We construct the sketch by the following procedure.
\begin{enumerate}[leftmargin=*]
    \item Initialize $u=v=0$.
    \item At the beginning of round $t$, record $v_t\gets v$.
    While $|h_t(v)|>\eps/8$, update
    \begin{align*}
        u\gets u-(\eps/32)\sign(h_t(v))x_t,\qquad
        v\gets\frac{u}{\max\{1,\|u\|\}}.
        \numberthis\label{eq:OSupdate}
    \end{align*}
    Once $|h_t(v)|\leq\eps/8$, proceed to round $t+1$.
\end{enumerate}
First, we bound the total number of updates, including repetitions within a round. Index updates by $i$, let $t_i$ be their round indices, and let $v_i^{\textup{upd}}$ denote $v$ immediately before update $i$. Apply Lemma~\ref{lem:LazyOGD} to the convex losses
$\ell_i(v)\coloneq|h_{t_i}(v)|$ and step size $\beta=\eps/32$.
The subgradients have norm at most $1$ because $\|x_{t_i}\|\leq1$.
Moreover, $\ell_i(w)=0$ and $\ell_i(v_i^{\textup{upd}})>\eps/8$. For every positive finite prefix of $M$ updates,
\begin{align*}
    M\eps/8
    <\sum_{i=1}^M\ell_i(v_i^{\textup{upd}})
    \leq (2\beta)^{-1}+2\beta M
    =16/\eps+(\eps/16)M.
\end{align*}
Hence $M<256\eps^{-2}$. An infinite loop would have arbitrarily long finite prefixes, contradicting this bound. The procedure therefore terminates.

Let $S$ be the set of rounds containing at least one update. It serves both as the index set of the expert and as the exceptional set $D$ of the lemma, and $|S|\leq\min\{T,M_\eps\}$. For $t\notin D$, the recorded vector already satisfies $|h_t(v_t)|\leq\eps/8$. Summing updates on each context gives
\begin{align*}
    \beta_\tau\coloneq
    -(\eps/32)\sum_{i:t_i=\tau}\sign(h_\tau(v_i^{\textup{upd}}))
    \in(\eps/32)\zz.
\end{align*}
There are at most $M$ nonzero update steps, so $|\beta_\tau|\leq M\eps/32\leq8/\eps$.
Since $v_t$ was recorded before the current updates, unrolling yields
\begin{align*}
    v_t=
    \frac{\sum_{\tau\in S\cap[t-1]}\beta_\tau x_\tau}
    {\max\cbra{1,\left\|\sum_{\tau\in S\cap[t-1]}\beta_\tau x_\tau\right\|}}.
\end{align*}
Thus this is exactly $v_t(z;x_{<t})$ for the index set $S$ and coefficients $(\beta_\tau)_{\tau\in S}$ in~\eqref{eq:Zset}. The bias component does not affect reconstruction and can be chosen arbitrarily in $J_\eps$. This proves the lemma.
\end{proofof}

\subsection{Proof of Lemma~\ref{lem:SparseUpdateNash}}
In this section, we prove Lemma~\ref{lem:SparseUpdateNash}, the core result from \S\ref{sec:multibuyer}, which states that buyers report their values \emph{approximately truthfully in any Nash equilibrium}. For convenience, we restate Lemma~\ref{lem:SparseUpdateNash} below.
\vspace{1em}
\lemSparseUpdateNash*
\begin{proofof}{Lemma~\ref{lem:SparseUpdateNash}.}\label{proof:SparseUpdateNash}
Let $\B\in\nash(\A^\L_\sm,\E,\T,\gamma)$ and write $H\coloneq\bargam/(1-\bargam)$.
Suppose there are $i\in[n]$ and $t^\star\in\T_i$ such that the event
\begin{align*}
    G\coloneq\{|\theta_{t^\star}-b_{t^\star}|>\del_h\}
\end{align*}
has positive probability. We construct a profitable deviation.

\paragraph{Modified Strategy.}
Up to $t^\star$, buyer $i$ runs its original algorithm with its original private randomness. At $t^\star$ it first samples the bid that $\B_i$ would submit and computes its quantized bid. It can therefore determine $G$ using its own information, without observing the hidden current policy. On $G$, the deviating algorithm $\B'_i$ submits the truthful bid $\theta_{t^\star}$; off $G$, it submits the original bid.

For its continuation, $\B'_i$ keeps the virtual private history $\widehat{\mathcal H}_{i,t}$ introduced in \S\ref{sec:multibuyer}, with $t$ denoting the last completed round. It records the originally sampled bid at $t^\star$, and, once the price is revealed, computes the allocation and utility that this bid, after quantization, would have produced. At each later round $t\in\T_i$, it runs $\B_i$ using the virtual private history $\widehat{\mathcal H}_{i,t-1}$ together with the current context and value. After each later round, it appends its actual new observations, together with the bids and private random variables generated by $\B_i$, to the record. This is a valid strategy: the buyer knows the original bid, the revealed price, and its value.

\paragraph{Coupled Protocols.}
Couple the original and deviating runs using the same seller and buyer random seeds. Each fresh seller draw uses an independent seed, and the coupling uses identical draws whenever the relevant histories agree. Use primes for the deviating run. The two runs agree through the original bid calculation at $t^\star$. Consequently,
$x_{t^\star}=x'_{t^\star}$, $\theta_{t^\star}=\theta'_{t^\star}$, and $p_{t^\star}=p'_{t^\star}$.
On $G$, the deviating quantized bid is $b'_{t^\star}=Q_h(\theta_{t^\star})$.
Off $G$, both runs agree throughout.

\paragraph{Utility Decomposition.}
The past utility is identical. Writing $\Delta u_t$ for the deviating minus original per-round utility, the coupled discounted utility difference is
\begin{align*}
    \sum_{t\in\T_i}\gamma_i^{t-1}\Delta u_t
    =\ind[G]\bigg(
       \gamma_i^{t^\star-1}\Delta u_{t^\star}
       +\sum_{\substack{t\in\T_i\\t>t^\star}}
             \gamma_i^{t-1}\Delta u_t
      \bigg).
\end{align*}

\paragraph{Effect of Sparse Update.}
If $\xi_{t^\star}=0$, the changed bid is not fed to $\L$, and the seller retains no other pricing state that depends on bids, so its future expert functions depend only on the public contexts. Neither the other buyers nor $\E$ observe the changed bid or allocation, and buyer $i$ continues to run its original algorithm on the virtual private history $\widehat{\mathcal H}_{i,t}$. Induction on the subsequent rounds therefore couples all future public observations, bids, and utilities identically on this event, for every permitted environment in the sense of \S\ref{sec:prelim}.

When $\xi_{t^\star}=1$, we bound the future difference directly. A buyer pays only when the price is at most its quantized bid, hence at most $1$, so its utility in one round lies in $[-1,1]$ even if it overbids. Allowing the future values in the two runs to differ, we obtain
\begin{align*}
    \sum_{\substack{t\in\T_i\\t>t^\star}}
        \gamma_i^{t-1}\Delta u_t
    \geq
    -2\ind[\xi_{t^\star}=1]\,
         \frac{\gamma_i^{t^\star}}{1-\gamma_i}
    \geq
    -2\ind[\xi_{t^\star}=1]\gamma_i^{t^\star-1}H.
    \numberthis\label{eq:variableststar}
\end{align*}

\paragraph{Conditioning over Random Pricing.}
Condition on all past randomness, the current context and value, and the originally sampled bid, but not on the fresh current seller draws. Conditional on the past, these draws are independent of the current context, value, and bid: the current pair is selected without seeing the current private policy, and the bid is submitted before that policy is revealed. On exploration, Lemma~\ref{lem:RandomPricing} gives the current gain from $Q_h(\theta_{t^\star})$. On exploitation, that quantized truthful bid loses at most $h$ relative to $u(\theta_{t^\star},p_{t^\star},\theta_{t^\star})$, the maximum current utility before quantization. Therefore, on $G$,
\begin{align*}
    \expec[\Delta u_{t^\star}\mid\textup{conditioned information}]
    &\geq
    \frac{\eps}{2(1+h)}
       [|\theta_{t^\star}-b_{t^\star}|-2h]_+^2-h\\
    &\geq
    \frac{\eps}{4}
       [|\theta_{t^\star}-b_{t^\star}|-2h]_+^2-h\\
    &>
    \frac{\eps}{4}(\del_h-2h)^2-h.
\end{align*}
Here $h\leq1$ and $G$ implies $|\theta_{t^\star}-b_{t^\star}|>\del_h>2h$.
The update coin is fresh and independent, so its conditional probability is $\rho$.

\paragraph{Taking Expectation.}
Combining the current gain with~\eqref{eq:variableststar} and taking marginal expectations yields
\begin{align*}
    &U_i(\B'_i,\B_{-i};\A^\L_\sm,\E,\T)
      -U_i(\B_i,\B_{-i};\A^\L_\sm,\E,\T)\\
    &\qquad\geq
    \Pr(G)\gamma_i^{t^\star-1}
       \bigg(\frac{\eps}{4}(\del_h-2h)^2-h-2\rho H\bigg)\\
    &\qquad=
    \Pr(G)\gamma_i^{t^\star-1}(h+2\rho H)>0.
\end{align*}
The equality uses~\eqref{eq:neartruthradius}. Positivity follows from $\Pr(G)>0$, $\gamma_i>0$, and $h>0$. This contradicts Nash equilibrium.

Finally, the choices~\eqref{eq:gridprecision} and~\eqref{eq:updateprob} give $h\leq\eps^3/1024$ and $\rho H\leq\eps^3/2048$. Thus
\begin{align*}
    \del_h
    \leq \eps^3/512+\eps/8
    <\eps/4
\end{align*}
for $\eps\in(0,1/4]$. Taking a union over the finitely many rounds shows that the conclusion holds almost surely.
\end{proofof}

\subsection{Existence of an Equilibrium for the Finite Mechanism}\label{app:finitegame}
Fix a deterministic permitted environment and finite $T$. Each quantized bid belongs to the finite set $\G_h$. Each seller draw has finite support: an expert, an exploration grid point, or a Bernoulli coin. Given a finite history and these choices, the next context, value, price, and allocation are determined. Induction on the horizon therefore gives a finite game tree, even though the contexts and prices themselves may be real numbers. This finite extensive-form game has a mixed Nash equilibrium.

Implement each quantized bid $b\in\G_h$ by submitting the bid $\widetilde b=b$. Any deviation with bids in $[0,1]$ induces a distribution over contingent plans for quantized bids in the finite tree. The seller's pricing state depends on a submitted bid only through its quantized value, and the bid is private to the seller and the buyer. A buyer that remembers its exact bid within the quantization cell therefore gains only private randomization, which mixed strategies already allow. Thus a profitable deviation with bids in $[0,1]$ would induce a profitable deviation in the finite game, proving that its equilibrium lifts to the game with bids in $[0,1]$.

\section{Miscellaneous Proofs}\label{app:miscproof}
In this section, we provide proofs of technical lemmas used in the main results.
\vspace{1em}
\lemZapprox*
\begin{proofof}{Lemma~\ref{lem:Zapprox}.}\label{proof:Zapprox}
Fix $f_{w,a}\in\F_{\textup{aff}}$. If $a\leq-1$, its price is zero on $\bb^d$; if $a>2$, its price exceeds $1$ and it never sells. In either case its revenue is zero, as is the revenue of the expert with index set $S=\emptyset$ and bias component $-1$. For $a\in[-1,2]$, take the index set $S$, the coefficients $(\beta_\tau)_{\tau\in S}$, and the exceptional set $D$ from Lemma~\ref{lem:OnlineSketch} applied to $w$, and let $a(z)$ be the largest grid point in $J_\eps$ not exceeding $a$. Since the grid spacing is at most $\eps/8$, for every $t\notin D$,
\begin{align*}
    -\eps/4
    \leq \innp{v_t(z;x_{<t}),x_t}+a(z)
          -\innp{w,x_t}-a
    \leq\eps/8.
    \numberthis\label{eq:Zapprox-1}
\end{align*}
Let $p=f_{w,a}(x_t)$ and $q=f_t(z;x_t)$. If $p>0$, subtracting $\eps/2$ and taking the positive part gives
\begin{align*}
    [p-3\eps/4]_+\leq q\leq[p-3\eps/8]_+\leq p.
    \numberthis\label{eq:Zpparox-2}
\end{align*}
If the comparator sells, then $b_t\geq p\geq q$, so the expert also sells and loses at most $3\eps/4$ in revenue. If the comparator does not sell, or $p=0$, its revenue is zero and the same revenue bound holds by nonnegativity. Thus
\begin{align*}
    \irev(f_{w,a},x_t,b_t)
    \leq\irev(f_t(z;\cdot),x_t,b_t)+3\eps/4
    \qquad(t\notin D).
    \numberthis\label{eq:Zapprox-3}
\end{align*}
On $D$, charge at most $1$ per round. Summing yields
\begin{align*}
    \sum_{t=1}^T\irev(f_{w,a},x_t,b_t)
    \leq
    \sum_{t=1}^T\irev(f_t(z;\cdot),x_t,b_t)
    +(3\eps/4)T+\min\{T,M_\eps\}.
\end{align*}
Taking the supremum over comparators proves the lemma.
\end{proofof}
\lemRevStab*
\begin{proofof}{Lemma~\ref{lem:RevStab}.}\label{proof:RevStab}
Fix $f=f_{w,a}$ and define $f'=f_{w,a-\del}\in\F_{\textup{aff}}$. Pointwise,
$f'(x)=[f(x)-\del]_+$.
If $f(x_t)\leq\theta_t$, then
$f'(x_t)\leq[\theta_t-\del]_+\leq b_t$, so the shifted policy sells and
\begin{align*}
    \irev(f,x_t,\theta_t)
    =f(x_t)\leq f'(x_t)+\del
    =\irev(f',x_t,b_t)+\del.
\end{align*}
If the comparator does not sell, its revenue is zero and this inequality still holds. Summing and taking the supremum over $f$ proves the result.
\end{proofof}
\lemRandomPricing*
\begin{proofof}{Lemma~\ref{lem:RandomPricing}.}\label{proof:RandomPricing}
Write $q=Q_h(\theta)$ and
$L\coloneq\expec_{\lambda\sim U_{\G_h}}[u(\theta,\lambda,q)-u(\theta,\lambda,b)]$.
Since both $q$ and $b$ are grid points, direct summation gives
\begin{align*}
    L=
    \begin{cases}
        \displaystyle\frac{h}{1+h}
        \sum_{\substack{p\in\G_h\\b<p\leq q}}(\theta-p),&b\leq q,\\[2mm]
        \displaystyle\frac{h}{1+h}
        \sum_{\substack{p\in\G_h\\q<p\leq b}}(p-\theta),&b>q.
    \end{cases}
\end{align*}
Here a grid point has probability $1/(m+1)=h/(1+h)$.
If $b=q$, then $L=0$ and $|\theta-b|=\theta-Q_h(\theta)<h$, so the right side of the claimed bound is zero and the bound holds.

Otherwise let $k=|b-q|/h\geq1$. In the underbid case, the $k$ summands are at least $0,h,\dots,(k-1)h$, in reverse order. In the overbid case the same lower bounds follow from $q\leq\theta<q+h$. This latter case cannot arise when $q=1$. Therefore,
\begin{align*}
    L\geq\frac{h^2k(k-1)}{2(1+h)}
     \geq\frac{h^2(k-1)^2}{2(1+h)}
     \geq\frac{[|\theta-b|-2h]_+^2}{2(1+h)},
\end{align*}
where the last step uses $|\theta-b|\leq(k+1)h$.

Finally, $u(\theta,p,\theta)$ maximizes $u(\theta,p,b)$ over $b\in[0,1]$ at every fixed price $p\geq0$. Replacing $\theta$ by the quantized truthful bid $q$ changes the allocation only if $q<p\leq\theta$; the forgone utility is then $\theta-p<h$. This proves the last assertion, including prices above $1$.
\end{proofof}

\subsection{IID Online-to-Batch Conversion}\label{app:onlinetobatch}
Let $(X,\Theta)$ have the IID data distribution $P$, and let
$\rev_P(f)\coloneq\expec[f(X)\ind[\Theta\geq f(X)]]$.
Since the policy $f_t$ is selected before the fresh pair $(x_t,\theta_t)$ is drawn,
$\expec[\irev(f_t,x_t,\theta_t)]=\expec[\rev_P(f_t)]$.
Moreover, $Q_h(\theta_t)\leq\theta_t$ gives
$\irev(f_t,x_t,b_t)\leq\irev(f_t,x_t,\theta_t)$.
Using the expected regret definition,
\begin{align*}
    T\sup_{f\in\F_{\textup{aff}}}\rev_P(f)
    &\leq
    \expec\sbra{\sup_{f\in\F_{\textup{aff}}}
        \sum_{t=1}^T\irev(f,x_t,\theta_t)}\\
    &=\reg(\A^\L_\omr;\E,\T)
       +\sum_{t=1}^T\expec[\irev(f_t,x_t,b_t)]\\
    &\leq\reg(\A^\L_\omr;\E,\T)
       +\sum_{t=1}^T\expec[\rev_P(f_t)].
\end{align*}
For an independent $J\sim U_{[T]}$ chosen once after training,
$\expec[\rev_P(f_J)]=T^{-1}\sum_t\expec[\rev_P(f_t)]$.
Dividing the preceding bound by $T$ proves the claim. This is an expected guarantee for a single randomly selected policy $f_J\in\F_{\textup{aff}}$.

\section{Proof of Theorem~\ref{thm:SRegretOMR}}\label{app:thm:SregretOMR}
In this section, we present the full proof of Theorem~\ref{thm:SRegretOMR}, restated below for convenience:
\vspace{1em}
\thmSRegretOMR*
\begin{proofof}{Theorem~\ref{thm:SRegretOMR}.}
Fix $\B\in\nash(\A^\L_\sm,\E,\T,\gamma)$. By Lemma~\ref{lem:SparseUpdateNash},
$|\theta_t-b_t|\leq\del_h<\eps/4$ for every $t$, almost surely.
Lemma~\ref{lem:RevStab} therefore gives
\begin{align*}
    &\expec_{\omrgameSR}\sbra{
        \sup_{f\in\F_{\textup{aff}}}\sum_{t=1}^T\irev(f,x_t,\theta_t)
        -\sum_{t=1}^T\irev(f_t,x_t,b_t)}\\
    &\quad\leq
    \expec_{\omrgameSR}\sbra{
        \sup_{f\in\F_{\textup{aff}}}\sum_{t=1}^T\irev(f,x_t,b_t)
        -\sum_{t=1}^T\irev(f_t,x_t,b_t)}
    +\del_hT.
    \numberthis\label{eq:bid-stability-applied-to-regret}
\end{align*}
Here $f_t$ denotes the policy actually used in round $t$, including on exploration rounds.

For the analysis, on exploration rounds we also draw an unused $z_t$ from the current oracle distribution, using a fresh private seed before the context is observed. This does not change the mechanism. Let
$r_t(z)=\irev(f_t(z;\cdot),x_t,b_t)$.
The actual revenue is at least $r_t(z_t)-\ind[\omega_t=1]$, since both revenues lie in $[0,1]$ and they agree on exploitation rounds. Lemma~\ref{lem:Zapprox} and $\expec[\sum_t\ind[\omega_t=1]]=\eps T$ imply
\begin{align*}
    &\expec_{\omrgameSR}\sbra{
        \sup_{f\in\F_{\textup{aff}}}\sum_{t=1}^T\irev(f,x_t,b_t)
        -\sum_{t=1}^T\irev(f_t,x_t,b_t)}\\
    &\quad\leq
    \expec\sbra{\sup_{z\in\Z_{\eps,T}}
         \sum_{t=1}^T(r_t(z)-r_t(z_t))}
    +(3\eps/4)T+\min\{T,M_\eps\}+\eps T\\
    &\quad\leq
    \rho^{-1}R(\rho T\log|\Z_{\eps,T}|)
    +\O\big(\sqrt{\rho^{-1}T\log(2|\Z_{\eps,T}|T)}\big)
    +(7\eps/4)T+\min\{T,M_\eps\}.
    \numberthis\label{eq:thm5-almost-there}
\end{align*}
The last step uses Lemma~\ref{lem:SparseUpdateRegret}, whose conditions hold here: the environment is non-anticipating because the current bid is sealed before the fresh private policy is revealed and the current pair does not depend on that policy, and the update coin is generated after the bid is submitted.

Combining~\eqref{eq:bid-stability-applied-to-regret} and~\eqref{eq:thm5-almost-there}, then taking the supremum over $\B$, proves the generic bound.
For Hedge, $R(x)=\O(\sqrt{x})$,
$\log|\Z_{\eps,T}|=\O(\eps^{-2}\log(T/\eps))$, and
\begin{align*}
    \rho^{-1}
    \leq1+\frac{2048\bargam}{(1-\bargam)\eps^3}.
\end{align*}
Also $\min\{T,M_\eps\}\leq\sqrt{TM_\eps}$.
Substituting gives the stated rate.
Sampling from the grid only requires a uniform integer from $\{0,\dots,m\}$, so the running time is dominated by expert enumeration, as in Theorem~\ref{thm:RegretOMR}.
\end{proofof}

\subsection{Regret under Sparse Feedback}
In this section, we prove the following lemma on the regret overhead introduced by sparse updating. The proof is a direct application of Bernstein's inequality for martingales, stated in Lemma~\ref{lem:BernsteinMartingale}.
\vspace{1em}
\begin{restatable}{lemmma}{lemSparseUpdateRegret}
    Let $\L$ be an expert oracle over $K$ experts, whose state depends only on the supplied reward vectors, with fresh independent private expert draws. Suppose its expected regret after $n$ rounds against every non-anticipating adaptive environment is at most $R(n\log K)$, where $R:[0,\infty)\to[0,\infty)$ is nonnegative, nondecreasing, and concave.
    Fix $\rho\in(0,1]$, and let $\L_\rho$ be the following procedure. In each round $t$, it draws
    \begin{align*}
        z_t\sim\L((r_\tau)_{\tau<t:\xi_\tau=1}),
    \end{align*}
    and, after $r_t\in[0,1]^K$ is fixed, it draws $\xi_t\sim\ber(\rho)$ independently and feeds $r_t$ to $\L$ only if $\xi_t=1$. The current reward vector is fixed without observing the fresh private draw $z_t$, and neither the oracle nor the environment uses future update coins. Then
    \begin{align*}
        \reg^{\textup{exp}}(\L_\rho;\E)
        \leq\rho^{-1}R(\rho T\log K)
        +\O\big(\sqrt{\rho^{-1}T\log(2KT)}\big).
    \end{align*}
    \label{lem:SparseUpdateRegret}
\end{restatable}

\begin{proofof}{Lemma~\ref{lem:SparseUpdateRegret}.}\label{proof:SparseUpdateRegret}
The case $K=1$ has zero regret, so assume $K\geq2$.
Let $\F_{t-1}$ contain all information immediately before $\xi_t$ is drawn, including past randomness, current external observations, $r_t$, and $z_t$. Let $\F_T$ contain the completed interaction. These sigma-algebras are nested: $\F_t$ also contains the randomness and observations between $\xi_t$ and the next update coin. The fresh coin $\xi_t$ is independent of $\F_{t-1}$.

Fix $z\in[K]$, and define
\begin{align*}
    X_t(z)\coloneq r_t(z)-r_t(z_t),\qquad
    Y_t(z)\coloneq\xi_tX_t(z).
\end{align*}
Then $X_t(z)$ is $\F_{t-1}$-measurable, $|X_t(z)|\leq1$, and
\begin{align*}
    \expec[Y_t(z)\mid\F_{t-1}]=\rho X_t(z).
    \numberthis\label{eq:Y_t_expec}
\end{align*}
Thus $Z_t(z)\coloneq(\rho-\xi_t)X_t(z)$ is a martingale difference, with
$|Z_t(z)|\leq2$ and
\begin{align*}
    \expec[Z_t(z)^2\mid\F_{t-1}]
    =\rho(1-\rho)X_t(z)^2\leq\rho.
    \numberthis\label{eq:cond_var_Z_t}
\end{align*}
Lemma~\ref{lem:BernsteinMartingale}, followed by a union bound over $K$ experts, gives, for all $s\geq0$,
\begin{align*}
    \Pr\sbra{
       \max_{z\in[K]}\sum_{t=1}^T Z_t(z)
       >\sqrt{2\rho T(\log K+s)}
           +\frac{2\sqrt{2}}{3}(\log K+s)}
    \leq e^{-s}.
\end{align*}
Integrating this tail, using
$\sqrt{\log K+s}\leq\sqrt{\log K}+\sqrt{s}$, yields
\begin{align*}
    \expec\sbra{\sbra{\max_{z\in[K]}
         \sum_{t=1}^T Z_t(z)}_+}
    \leq\O\big(\sqrt{\rho T\log(2K)}+\log(2K)\big).
\end{align*}
Consequently,
\begin{align*}
    \reg^{\textup{exp}}(\L_\rho;\E)
    &\leq
    \rho^{-1}\expec\sbra{\sup_{z\in[K]}
       \sum_{t=1}^T\xi_t(r_t(z)-r_t(z_t))}\\
    &\quad+\O\big(\sqrt{\rho^{-1}T\log(2K)}
                        +\rho^{-1}\log(2K)\big).
    \numberthis\label{eq:regexp-pan-ultimate-bound}
\end{align*}

To bound the first term, we may equivalently generate the entire coin sequence independently at the beginning, as long as no party inspects future coins. Condition on this sequence; its selected times are then fixed, with $N=\sum_t\xi_t$. The random seeds used to sample experts and the external random seeds remain independent of these coins. At each selected time, the oracle receives exactly the earlier selected rewards, and the reward vector cannot depend on its current fresh draw. Draws at unselected times can affect later rewards and are part of the intervening adaptive environment; they do not change the oracle's reward history. Thus, for each fixed schedule, the selected interaction is a valid non-anticipating environment of $N$ rounds for $\L$. The assumed regret bound for each fixed number of rounds gives
\begin{align*}
    \expec\sbra{\sup_{z\in[K]}
       \sum_{t=1}^T\xi_t(r_t(z)-r_t(z_t))
       \,\middle|\,\xi_{\leq T}}
    \leq R(N\log K).
\end{align*}
If $N=0$, the sum on the left is empty and the bound holds because $R(0)\geq0$. The conditioning fixes only the update schedule; the remaining independent seeds retain their original laws.
By concavity,
\begin{align*}
    \expec\sbra{\sup_{z\in[K]}
       \sum_{t=1}^T\xi_t(r_t(z)-r_t(z_t))}
    \leq\expec[R(N\log K)]
    \leq R(\rho T\log K).
\end{align*}

Finally, put $L=\log(2KT)$. If $\rho T\geq L$, the linear term $\rho^{-1}\log(2K)$ in~\eqref{eq:regexp-pan-ultimate-bound} is at most
$\sqrt{\rho^{-1}TL}$. If $\rho T<L$, the trivial regret bound $T$ is already at most $\sqrt{\rho^{-1}TL}$. In both cases, nonnegativity of $R$ gives the claimed bound.
\end{proofof}
\begin{restatable}{lemmma}{lemBernsteinMartingale}\textup{[Bernstein's inequality for martingales, Lemma A.8 in~\cite{cesa2006prediction}]}
    Let $Z_1,\dots,Z_T$ be a bounded martingale difference sequence with respect to the filtration $(\F_t)_{t=0}^T$, and with $|Z_t|\leq c$. Then, for all constants $\theta, \nu>0$,
    \begin{align*}
        \Pr\sbra{\,\sum_{t=1}^T Z_t>\sqrt{2\nu \theta}+\big(\sqrt{2}/3\big)c\theta\ \ \textup{and}\ \ \Sigma^2_T\leq \nu\,}\leq e^{-\theta},
    \end{align*}
    where $\Sigma^2_T\coloneq \sum_{t=1}^T\expec[Z_t^2|\F_{t-1}]$ is the sum of the conditional variances.
    \label{lem:BernsteinMartingale}
\end{restatable}

\section{Proof of Lazy OGD Regret Bound}\label{app:lazyOGD}
In this section, we prove Lemma~\ref{lem:LazyOGD}, which establishes the regret bound for Lazy Online Gradient Descent. This result follows from the regret bound for the lazy variant of Online Mirror Descent when using the squared Euclidean distance $\frac{1}{2}\|x- y\|^2$ as the Bregman divergence; see, for example, \citet{bubeck2015convex,hazan2016introduction}. Here, we present a more direct proof.
\vspace{1em}
\lemLazyOGD*
\begin{proofof}{Lemma~\ref{lem:LazyOGD}}
    First, unroll the update rule for $u_i$ to write
    \begin{align*}
        u_i = - \beta \sum_{s=1}^{i-1} \nabla f_s(v_s),
    \end{align*}
    since $u_1=0$. Let $L_i(x)\coloneq \frac{1}{2}\|x\|^2+ \beta\sum_{s=1}^{i-1}\big\langle\nabla f_s(v_s),x\big\rangle$. Then, we can rewrite $v_i$ as
    \begin{align*}
        v_i\coloneq\frac{u_i}{\max(1,\|u_i\|)}=\argmin_{x \in \bb^d}\|x - u_i\|^2&= \argmin_{x \in \bb^d}\Big(\|x\|^2 - 2\innp{u_i,x}\Big)\\
        &= \argmin_{x \in \bb^d}\Big(\bunderbrace{\|x\|^2 + 2\beta\sum_{s=1}^{i-1}\innp{\nabla f_s(v_s),x}}{=2L_i(x)}\Big)\\
        &=\argmin_{x \in \bb^d}L_i(x).
    \end{align*}

    Since $L_i$ is $1$-strongly convex with respect to $\|\cdot\|$,
    \begin{align*}
        L_{i+1}(v_{i+1}) - L_{i+1}(v_i) \leq \big\langle \nabla L_{i+1}(v_{i+1}), v_{i+1} - v_i \big\rangle - \frac{1}{2} \|v_{i+1} - v_i\|^2 \leq -\frac{1}{2} \|v_{i+1} - v_i\|^2,
    \end{align*}
    where the last inequality follows from the optimality of $v_{i+1}=\argmin_{x \in \bb^d} L_{i+1}(x)$ and the convexity of $L_{i+1}$. On the other hand,
    \begin{align*}
        L_{i+1}(v_{i+1})-L_{i+1}(v_i)\geq  L_i(v_{i+1})-L_i(v_i)+\beta\big\langle \nabla f_i(v_i), v_{i+1}-v_i\big\rangle \geq \beta \big\langle \nabla f_i(v_i), v_{i+1}-v_i\big\rangle,
    \end{align*}
    as $v_i=\argmin_{x\in \bb^d}L_i(x)$. Combining the above inequalities,
    \begin{align*}
        \frac{1}{2}\|v_{i+1}-v_i\|^2 \leq -\beta\big\langle \nabla f_i(v_i), v_{i+1}-v_i\big\rangle \leq \beta \|\nabla f_i(v_i) \| \|v_{i+1}-v_i\|,
    \end{align*}
    hence $\|v_{i+1}-v_i\| \leq 2\beta \|\nabla f_i(v_i)\|$ and
    \begin{align*}
        \big\langle \nabla f_i(v_i), v_i-v_{i+1}\big\rangle \leq \|\nabla f_i(v_i)\| \,\| v_{i+1}-v_i \|\leq 2\beta \|\nabla f_i(v_i)\|^2.
        \numberthis\label{eq:ogd-inequality-1}
    \end{align*}

    Next, we claim the following by induction: For all $i \leq M$, we have
    \begin{align*}
        \forall x \in \bb^d,\quad\sum_{s=1}^{i} \big\langle  \nabla f_s(v_s),v_{s+1}-x\big\rangle \leq \frac{\|x\|^2}{2\beta}.
        \numberthis\label{eq:ogd-inequality-2}
    \end{align*}
    For $i=0$, the above holds as $0\leq \|x\|^2/2\beta$. Suppose \eqref{eq:ogd-inequality-2} holds for some $i\leq M-1$. Then, it holds in particular for $x=v_{i+2}\coloneq\argmin_{x\in \bb^d} L_{i+2}(x)$ and
    \begin{align*}
        \sum_{s=1}^{i+1} \big\langle \nabla f_s(v_s),v_{s+1} \big\rangle &= \big\langle \nabla f_{i+1}(v_{i+1}),v_{i+2} \big\rangle +  \sum_{s=1}^{i} \big\langle \nabla f_s(v_s),v_{s+1}\big\rangle \\
        &\leq \big\langle \nabla f_{i+1}(v_{i+1}),v_{i+2} \big\rangle + \sum_{s=1}^{i} \big\langle \nabla f_s(v_s),v_{i+2}\big\rangle + \frac{\|v_{i+2}\|^2}{2\beta} \tag{by the induction hypothesis}\\
        &= \bunderbrace{\sum_{s=1}^{i+1} \big\langle \nabla f_s(v_s),v_{i+2}\big\rangle + \frac{\|v_{i+2}\|^2}{2\beta}}{=\beta^{-1}L_{i+2}(v_{i+2})}\\
        &\leq \bunderbrace{\sum_{s=1}^{i+1} \big\langle \nabla f_s(v_s),x\big\rangle + \frac{\|x\|^2}{2\beta}}{=\beta^{-1}L_{i+2}(x)},\quad \forall x\in \bb^d. \tag{as $L_{i+2}(v_{i+2})\leq L_{i+2}(x)$}
    \end{align*}
    Hence, \eqref{eq:ogd-inequality-2} holds for $i+1$.

    Finally, combining~\eqref{eq:ogd-inequality-1} and~\eqref{eq:ogd-inequality-2} with $i=M$ and $x=v^\star$, we obtain
    \begin{align*}
        \sum_{i=1}^M\big[f_i(v_i)-f_i(v^\star) \big] &\leq \sum_{i=1}^M \big\langle \nabla f_i(v_i),v_i-v^\star \big\rangle \tag{as $f_i$ is convex.}\\
        &=\sum_{i=1}^M \big\langle \nabla f_i(v_i),v_{i+1}-v^\star \big\rangle + \sum_{i=1}^M \big\langle \nabla f_i(v_i),v_i-v_{i+1} \big\rangle\\
        &\leq \frac{\|v^\star\|^2}{2\beta} + 2\beta \sum_{i=1}^M \|\nabla f_i(v_i) \|^2 \tag{by~\eqref{eq:ogd-inequality-1} and~\eqref{eq:ogd-inequality-2}}\\
        &\leq \frac{1}{2\beta} + 2\beta M \tag{as $\|v^\star\|, \|\nabla f_i(v_i)\|\leq 1$}
    \end{align*}
\end{proofof}

\section{Other Related Work}\label{app:relatedwork}
In this section, we expand on the related work to more clearly situate our contributions within the existing literature.

\paragraph{Dynamic Pricing.}
The seminal work on online learning for pricing is by~\citet{kleinberg2003value}, who studied repeated posted-price auctions under both stochastic and adversarial environments. In the stochastic setting, buyer valuations are drawn from an unknown distribution, while in the adversarial case, they may be chosen adaptively and adversarially. This work laid the foundation for a broad literature on dynamic pricing, often with bandit feedback (observing only whether the buyer purchases the good), including variants such as demand learning~\citep{lin2006dynamic, besbes2009dynamic, besbes2015surprising}. Subsequent research has extended these models to settings with limited supply~\citep{babaioff2012dynamic, yang2014dynamic}, strategic buyer behavior~\citep{amin2013learning, liu2018learning}, and contextual information~\citep{chen2022primal, luo2024distribution}. For a comprehensive overview of this literature, see the survey by~\citet{den2015dynamic}.

\paragraph{Contextual Pricing.}
Learning an optimal pricing policy in the presence of contextual information has been studied in the batch learning setting by~\citet{mohri2014learning, munoz2017revenue, shen2019learning}. These works provide algorithms with PAC-style guarantees under the assumption that context–value pairs are drawn IID from an unknown distribution. A more recent contribution by~\citet{liu2020myersonian} investigates the computational and sample complexity of learning optimal linear pricing policies, a task they refer to as Myersonian regression. They show that obtaining a fully polynomial-time approximation scheme (FPTAS) is impossible under standard complexity-theoretic assumptions.

\end{document}